\documentclass[10pt]{iopart}
\usepackage{epsfig,graphicx}
\usepackage{amsfonts}

\newcommand{\sigmav}{\mbox{\boldmath$\sigma$}}
\newcommand{\Omegav}{\mbox{\boldmath$\Omega$}}
\newcommand{\hv}{\mbox{\boldmath$h$}}
\newcommand{\Hv}{\mbox{\boldmath$H$}}

\newcommand{\thetav}{\mbox{\boldmath$\theta$}}

\newcommand{\bra}{\langle}
\newcommand{\ket}{\rangle}

\newcommand{\order}{{\mathcal O}}

\newcommand{\bnull}{{\mbox{\boldmath $0$}}}
\newcommand{\be}{\begin{equation}}
\newcommand{\ee}{\end{equation}}
\newcommand{\bd}{\begin{displaymath}}
\newcommand{\ed}{\end{displaymath}}
\newcommand{\vsp}{\vspace*{3mm}}

\newcommand{\bx}{\ensuremath{\mathbf{x}}}
\newcommand{\by}{\ensuremath{\mathbf{y}}}

\newcommand{\bsigma}{{\mbox{\boldmath $\sigma$}}}

\newcommand{\here}{\makebox(0,0)}

\begin{document}

\title[Analysis of processes on finitely connected graphs I: vertex covering]{Dynamical replica analysis of processes on finitely connected random graphs I: vertex covering}
\author{A Mozeika and ACC Coolen}
\address{
Department of Mathematics, King's College London\\ The Strand,
London WC2R 2LS, UK }

\pacs{02.50.Ey, 05.90.+m, 64.60.Cn, 89.20.Ff}

\ead{alexander.mozeika@kcl.ac.uk,ton.coolen@kcl.ac.uk}

%\date{\today}

%\maketitle

\begin{abstract}
We study the stochastic dynamics of Ising spin models with random bonds, interacting on finitely connected Poissonnian random graphs.  We use the dynamical replica method to derive closed dynamical equations for the joint spin-field probability distribution, and solve these within the replica symmetry ansatz. Although the theory is developed in a general setting, with a view to future applications in various other fields, in this paper we apply it mainly to the dynamics of the Glauber algorithm (extended with cooling schedules) when running on the so-called  vertex cover optimization problem. Our theoretical predictions are tested against both Monte Carlo simulations and known results from equilibrium studies. In contrast to previous dynamical analyses based on deriving closed equations for only a small numbers of scalar order parameters, the agreement between theory and experiment in the present study is nearly perfect.
\end{abstract}

\section{Introduction}\label{section:Intro}

The interest in studying finitely connected (FC) spin systems on random graphs,
as introduced in \cite{Viana} more than twenty years ago, has grown in recent years.
For this there appear to be at least two reasons.
Firstly, FC spin systems can be seen as an intermediate step between fully connected mean-field spin models \cite{BOOK} and finite-dimensional spin models. Although they are still of the mean field type in the sense that random site permutations are irrelevant in the mean field limit, the finite connectivity introduces notions of site neighborhood, distances, etc. This attractive property has drawn many into this field, and we have by now achieved a thorough understanding of the equilibrium behavior of FC spin systems \cite{Kanter,BetheSG1,Monasson,BetheSG,SoftSpins}. Secondly, many optimization and decision problems in theoretical computer science can be mapped into models of FC spin systems. This mapping allowed such problems to be studied with analytic methods of statistical mechanics, and has been very fruitful especially in the study of K-SAT \cite{Ksat1,Ksat2}, vertex covering \cite{VCising1}, and graph coloring \cite{ColoringRG1,ColoringRG2}.

 Although our understanding of the equilibrium properties of FC spin systems is now quite advanced, that of
 the non-equilibrium behaviour of such systems
 is, despite recent progress \cite{diluteDyn,StochDyn,ApprSch,ParDyn,DRTfc,Goos,LLmodel}, still relatively limited in comparison. Dynamical studies
 are generally harder, by definition, as they incorporate the equilibrium state as a special case. In the domain of the dynamics of FC spin systems
 the generating functional method (or path integration technique) of \cite{dD} is the only exact method available today. There has been some success in applying this method to finitely connected soft spin systems \cite{diluteDyn,StochDyn} and Ising spin systems \cite{ParDyn}. However,  the
 generating functional method
 leads one in FC systems to a formalism involving a rather complicated dynamical order parameter (describing the joint statistics of single-site
 spin `paths' and single site field perturbation `paths') which is generally difficult to handle. Even for parallel dynamics \cite{ParDyn} it is effectively
 equivalent to having a number of scalar order parameters that grows exponentially with the number of discrete time steps considered.
 For that reason, even in generating functional analysis studies one is in practice forced to make further approximations to tame this explosion of order parameters.

  An alternative approach to the dynamics of FC spin systems
 is dynamic replica theory (DRT) \cite{DRTinf1,DRTinf2}, which was initially developed for fully connected systems. In contrast to generating functional analysis, DRT in its present form is not (yet) exact; however, one can increase its accuracy systematically by increasing the size of the chosen order parameter set \cite{DRTinf2}.  The  great advantage of DRT in the study of FC spin systems, compared to generating functional analysis,  is that the effective number of order parameters does {\em not} grow with time.
Recently, the DRT method \cite{DRTfc} and its equivalent \cite{ApprSch} were used to study the
 dynamics of FC Ising spin systems, but only for a relatively small number of dynamic order parameters.
 Although its performance on regular random graphs was found to be very good \cite{DRTfc,ApprSch}, for random Poissonian graphs it was found to be quite poor \cite{DRTfc}. In the present  paper we develop the DRT method further, and cure the previous limitations by increasing the size of the order parameter set, following \cite{DRTinf2,ApprSch}, to the full joint spin-field distribution. We then  demonstrate the performance of the resulting improved theory by application to the so-called minimal vertex cover problem \cite{VCising1} on Poissonnian random graphs.

 This paper is organized as follows. In section \ref{section:Model} we define our model and derive an exact dynamical equation for the joint spin-field probability distribution. In the next section \ref{section:RA} we close this equation using the standard assumptions and procedures of DRT. We simplify our dynamical theory by making the standard replica symmetry ansatz in section \ref{section:RS}. In section \ref{section:VC} we apply our resulting formalism to
 the dynamics of the Glauber algorithm, extended with simulated annealing type cooling schedules, when running on the minimal vertex cover
  problem. The outcome of solving our dynamical equations numerically are compared to measurements taken in Monte Carlo simulations. Finally, in section \ref{section:Discussion} we summarize and discuss our results.
\section{Model definitions and  macroscopic laws}\label{section:Model}
\indent We  consider a system of $N$ Ising spins, $\sigma_i
\in\{-1,1\}$, which are placed on the vertices of a random Erd\"os-R\'enyi graph
\cite{Erdos}. Spins interact only when they are connected.
Their microscopic dynamics are governed by a Glauber
type stochastic algorithm. At each iteration of this algorithm a site
$i$ is drawn randomly from the set $\lbrace 1,\ldots, N\rbrace$ of
all sites, and spin $\sigma_i$ is subsequently flipped with probability
\begin{eqnarray}
\mathrm{P}(\sigma_i\rightarrow -\sigma_i)=\frac{1}{2}[1-
\sigma_i\tanh[\beta h_i(\sigmav)]]\label{eq:algorithm}
\end{eqnarray}
where $h_i$ is
a local field, defined as
\begin{eqnarray}
h_i(\sigmav)=\sum_{j\neq i} c_{ij}J_{ij}\sigma_j+
\theta\label{eq:field}
\end{eqnarray}
with $\sigmav=(\sigma_1,\ldots,\sigma_N)$. The inverse temperature
$\beta=T^{-1}$ controls the level of noise in the system; the
dynamics is random for $\beta=0$, and fully deterministic for
$\beta\rightarrow\infty$. The parameter $\theta$ defines a uniform
external field. The set of random variables $\{c_{ij}J_{ij}\}$ is
regarded as a quenched disorder. The bonds $J_{ij}$ are symmetric,
viz. $J_{ij}=J_{ji}$, and drawn independently from a probability
distribution $P(J)$. The independently distributed random
variables $c_{ij}\in \{0,1\}$ are the entries of
a symmetric adjacency matrix with zeroes on the main diagonal, defining the random graph.
 In this paper we consider finitely connected (FC) random
graphs of the Erd\"os-R\'enyi
\cite{Erdos}  type, where
\begin{eqnarray}
\forall i<j:&~~~& P(c_{ij})=\frac{c}{N}\delta_{c_{ij},1}+
(1-\frac{c}{N})\delta_{c_{ij},0}\label{eq:P(c)}
\end{eqnarray}
with $c=O(N^{0})$. In the $N\rightarrow\infty$ limit
the average number of connections per spin (or vertex) remains finite,
and the distribution of connectivities (or vertex degrees) is given
by a Poisson distribution with mean $c$:
\begin{eqnarray}
P_c(k)&=& c^k \rme^{-c}/k!
\label{eq:Poisson}
\end{eqnarray}
The process (\ref{eq:algorithm}) can be written in the form of a master
equation for the evolution of the microscopic state probability
in continuous time:\footnote{This involves formally the introduction
of random durations for the individual spin updates which are $N^{-1}$ on average, and which
for finite $N$ are drawn from a specific distribution \cite{Bedeaux}.}
\begin{eqnarray}
\frac{\mathrm{d}}{\mathrm{d}t}p_t(\sigmav) &=& \sum_{i=1}^N [p_t(F_i
\sigmav)w_i(F_i \sigmav)-
p_t(\sigmav)w_i(\sigmav)]\label{eq:master}
\end{eqnarray}
in which $F_i$ is a spin-flip operator $F_i \Omega (\sigmav)=\Omega
(\sigma_1,\ldots,-\sigma_i,\ldots,\sigma_N)$ and the quantities
$w_i(\sigmav)$ are the transition rates  given by
\begin{eqnarray}
w_i(\sigmav)=\frac{1}{2}[1-\sigma_i \tanh[\beta h_i(\sigmav)]]
\end{eqnarray}
This process evolves towards equilibrium Boltzmann probability
distribution $p_\infty (\sigmav)\sim \exp[-\beta H(\sigmav)]$,
with the Hamiltonian
\begin{eqnarray}
H(\sigmav)=-\sum_{i<j}\sigma_i
c_{ij}J_{ij}\sigma_j-\theta\sum_i \sigma_i
\end{eqnarray}
In general it is not possible to solve the $2^N$ coupled equations
(\ref{eq:master}) directly. Therefore, instead of following the
evolution of the microscopic distribution $p_t(\sigmav)$, one turns to
alternative descriptions of the dynamics in terms of macroscopic observables.
\vsp

For the reasons given in the introduction, we now follow the steps of dynamic replica theory \cite{DRTinf2},
 and consider the evolution in time of an arbitrary set of $\ell$ macroscopic
observables $\Omegav (\sigmav)=(\Omega_1
(\sigmav),\ldots,\Omega_\ell(\sigmav))$, where each
individual $\Omega_k(\sigmav)$ is taken to be of order $O(N^0)$.
We derive a Kramers-Moyal expansion for the associated macroscopic probability distribution
$P_t(\Omegav)=\sum_{\sigmav} \delta [\Omegav
-\Omegav(\sigmav)]p_t(\sigmav)$ by inserting the master equation
(\ref{eq:master}) into the time derivative of $P_t(\Omegav)$, and expanding the result in
powers of the 'discrete derivatives' $\Delta_i^\mu
(\sigmav)=\Omega_\mu(F_i\sigmav)- \Omega_\mu(\sigmav)$. This
gives:
\begin{eqnarray}
  \frac{\mathrm{d}}{\mathrm{d}t}P_t(\Omegav) & = & - \sum_{\mu=1}^\ell\frac{\partial}{\partial\Omega_{\mu}} \Big\{P_t(\Omegav)\Big\langle\sum_iw_i(\sigmav)\Delta_i^\mu
(\sigmav)\Big\rangle_{\Omegav ; t}\Big\}
\nonumber
\\
&&+\frac{1}{2}\sum_{\mu,\nu=1}^\ell\frac{\partial^2}{\partial\Omega_{\mu}\partial\Omega_{\nu}}
\Big\{P_t(\Omegav)\Big\langle\sum_iw_i(\sigmav)\Delta_i^\mu
(\sigmav)\Delta_i^\nu (\sigmav)\Big\rangle_{\Omegav ;
t}\Big\}\nonumber
\\
&&\hspace*{50mm} + O(N\ell^3 \Delta^3)
\label{eq:KM}
\end{eqnarray}
where we used the sub-shell (or conditional) average
\begin{eqnarray}
  \left\langle f(\sigmav)\right\rangle_{\Omegav ; t} & = &
  \frac{\sum_{\sigmav}p_t(\sigmav)\delta \left[\Omegav - \Omegav
(\sigmav)\right]f(\sigmav)}{\sum_{{\sigmav}}p_t({\sigmav})\delta
\left[\Omegav - \Omegav
  ({\sigmav})\right]}\label{eq:ssAv}
\end{eqnarray}
If the diffusion term in the expansion (\ref{eq:KM}) vanishes for
$N\rightarrow\infty$,  then (\ref{eq:KM}) acquires
the Liouville form, the solution of which describes the following deterministic flow:
\begin{eqnarray}
  \frac{\mathrm{d}}{\mathrm{d}t}\Omegav =
\Big\langle\sum_iw_i(\sigmav)\Big[\Omegav(F_i\sigmav)-\Omegav(\sigmav)\Big]\Big\rangle_{\Omegav
; t}
\label{eq:liouville}
\end{eqnarray}
This is then exact for
$N\rightarrow\infty$, but not necessarily  closed, due to the
presence of the microscopic probability $p_t(\sigmav)$ in (\ref{eq:ssAv}). In DRT, in order to close equation
(\ref{eq:liouville}),  one assumes equi-partitioning of probability
within the $\Omegav$ sub-shells, i.e. one takes $p_t(\sigmav)$ to depend on
$\sigmav$ only through \Omegav (\sigmav). The impact of this
assumption on the accuracy of the theory depends
critically on the choice of observables $\Omegav (\sigmav)$.

 In this paper, our choice of  observables $\Omegav
(\sigmav)$ is, as in  \cite{DRTinf2}, the (infinite dimensional) set given by the joint spin-field
distribution:
\begin{eqnarray}
D(s,h;\sigmav)=\frac{1}{N}\sum_i\delta_{s,\sigma_i}\delta\left[h-
h_i(\sigmav)\right] \label{eq:jsfield}
\end{eqnarray}
We assume that this
distribution (\ref{eq:jsfield}) is well behaved in the sense that it can be evaluated
first for a finite number $\ell$ of field arguments $h_{\mu}$, and that the limit $\ell\to\infty$
can be taken after the thermodynamic limit $N\to\infty$. For now on we thus have $2\ell$
observables $D(s,h_{\mu};\sigmav)$ with
$\mu=1,\ldots,\ell$ and $s\in\{-1,1\}$. In order to compute
 (\ref{eq:liouville}) we must work out the discrete derivatives
$\Delta_i^{s\mu}(\sigmav)=D(s,h_\mu;F_i\sigmav)-
D(s,h_\mu;\sigmav)$:
\begin{eqnarray}
\hspace*{-15mm}
  \Delta_i^{s\mu}(\sigmav) & = &
\frac{1}{N}\sum_j\delta_{s,F_i\sigma_j}\delta\left[h_{\mu}-
h_j(F_i\sigmav)\right]-\frac{1}{N}\sum_j\delta_{s,\sigma_j}\delta\left[h_{\mu}-
h_j(\sigmav)\right] \label{eq:discrder} \\
\hspace*{-15mm}
   & = & \frac{1}{N}\sum_{j\neq
i}\delta_{s,\sigma_j}c_{ij}\Big\{\delta_{\sigma_i,
1}\delta\left[h_{\mu}\!- h_j(\sigmav)\!+ 2J_{ij}\right]+
\delta_{\sigma_i,-1}\delta\left[h_{\mu}\!-
h_j(\sigmav)\!-2J_{ij}\right]
\nonumber\\[-2mm]
\hspace*{-15mm}
   &&\hspace*{23mm}-\delta\left[h_{\mu}\!- h_j(\sigmav)\right]\Big\}+
\frac{1}{N}\left\{\delta_{s,-\sigma_i}-\delta_{s,\sigma_i}\right\}\delta\left[h_{\mu}\!-
h_i(\sigmav)\right]\nonumber
\end{eqnarray}
Thus
$\Delta_i^{s\mu}(\sigmav)=O(N^{-1})$, so for $N\to\infty$ the diffusion term in
(\ref{eq:KM}) vanishes and the macroscopic observables $D(s,h_{\mu};\sigmav)$ evolve
deterministically according to (\ref{eq:liouville}). Inserting
(\ref{eq:discrder}) into (\ref{eq:liouville})
%and using general property $\sum_s\int\mathrm d h D(s,h;\sigmav)f(s,h)=\frac{1}{N}\sum_i f(\sigma_i,h_i(\sigmav))$ we
gives us a diffusion equation for the joint spin-field distribution:
\begin{eqnarray}
\hspace*{-15mm}
\frac{\partial}{\partial t}D(s,h_{\mu})&=&\frac{1}{2}\left [1+
s\tanh[\beta h_{\mu}]\right ]D(-
s,h_{\mu})-\frac{1}{2}\left [1- s\tanh[\beta
h_{\mu}]\right
]D(s,h_{\mu})\nonumber
\\
\hspace*{-15mm}
&&+\frac{1}{2}\sum_{{s^\prime}}\int\mathrm{d}{h^\prime}
[1\!-\!{s^\prime}\tanh[\beta
{h^\prime}]]\nonumber
\\
\hspace*{-15mm}&&\hspace*{10mm}\times
\Big\langle\frac{1}{N}\sum_{i\neq
j}\delta_{{s^\prime},\sigma_i}\delta_{s,\sigma_j}c_{ij}\delta [h^\prime\!\! -\! h_i (\sigmav)]\delta [ h_{\mu}\!\! -\! h_j
(\sigmav)\!+\! 2J_{ij}{s^\prime}]\Big\rangle_{D;t}\nonumber\\
\hspace*{-15mm}
&&-\frac{1}{2}\sum_{{s^\prime}}\int\mathrm{d}{h^\prime}
[1\!-\!{s^\prime}\tanh[\beta
{h^\prime}]]
\nonumber\\
\hspace*{-15mm}&&\hspace*{10mm}\times
\Big\langle\frac{1}{N}\sum_{i\neq
j}\delta_{{s^\prime},\sigma_i}\delta_{s,\sigma_j}c_{ij}\delta [
h^\prime\!\! -\! h_i (\sigmav)]\delta [ h_{\mu}\!\! -\! h_j
(\sigmav)]\Big\rangle_{D;t}
\label{eq:diffLong}
\end{eqnarray}
with the sub-shell average
\begin{eqnarray}
\left\langle f(\sigmav)\right\rangle_{D ; t} & = &
\frac{\sum_{\sigmav}p_t(\sigmav)f(\sigmav)\prod_{s\mu}\delta
\left[D(s,h_{\mu}) -
D(s,h_\mu;\sigmav)\right]}{\sum_{{\sigmav^\prime}}p_t({\sigmav^\prime})\prod_{s\mu}\delta
\left[D(s,h_{\mu}) -
D(s,h_\mu;{\sigmav^\prime})\right]}\label{eq:ssAvD}
\end{eqnarray}
The non-trivial objects in (\ref{eq:diffLong})
are the two averages, with angular brackets. To compute these efficiently we introduce the
following kernel, where $\tilde s \in \{0,s^\prime\}$,
\begin{eqnarray}
\hspace*{-15mm}
  \tilde{A}[s,s^\prime;h,h^\prime;\tilde s]&=&\Big\langle\frac{1}{c
  N}\sum_{i
j}\delta_{{s^\prime},\sigma_i}\delta_{s,\sigma_j}c_{ij}\delta [
h^\prime\!\! -\! h_i (\sigmav)]\delta [ h\! -\! h_j (\sigmav)\!+\!
2J_{ij}\tilde s]\Big\rangle_{D;t} \label{def:A}~~~~~~~
\end{eqnarray}
For $\tilde s =0$ the kernel (\ref{def:A}) defines the joint spin-field probability
of connected sites (a similar object was used to study the dynamics of the Ising
ferromagnet on a regular random graph \cite{ApprSch}).
In the limit $N\to\infty$, definition (\ref{def:A}) allows us to write
(\ref{eq:diffLong}) as
\begin{eqnarray}
\frac{\partial}{\partial t}D(s,h) &=& \frac{1}{2}\left [1+
s\tanh[\beta h]\right ]D(- s,h)-\frac{1}{2}\left [1-
s\tanh[\beta h]\right
]D(s,h) \nonumber
\\
&&+\frac{1}{2}c\sum_{{s^\prime}}\int\mathrm{d}{h^\prime}
[1-{s^\prime}\tanh[\beta {h^\prime}]]\tilde{A}[s,
s^\prime;h,
h^\prime;s^\prime]\nonumber\\
&&-\frac{1}{2}c\sum_{{s^\prime}}\int\mathrm{d}{h^\prime}
[1-{s^\prime}\tanh[\beta {h^\prime}]]\tilde{A}[s,
s^\prime;h, h^\prime;0].
\label{eq:diffusion}
\end{eqnarray}
%
%In the derivation of the above we anticipate that $l\rightarrow\infty$ limit will be taken, provided we let $N\rightarrow\infty$ first.
 This dynamical equation
(\ref{eq:diffusion}) is exact for large $N$, but not yet closed. Closure requires eliminating $p_t(\sigmav)$ from (\ref{def:A}).
\section{Replica analysis of the dynamics}\label{section:RA}
\subsection{Closure and disorder averaging}

To evaluate the right-hand side of (\ref{eq:diffusion}) we make the usual assumptions of the
 dynamic replica method. The observables $D(s,h_{\mu};\sigmav)$ are
taken to be self-averaging with respect to the disorder at any time, i.e. to
depend only on the {\em statistics} of the $\left\{c_{ij}J_{ij}\right\}$ rather than
their realization. Second, we assume equi-partitioning of the
microscopic probability within the $D(s,h_{\mu};\sigmav)$ sub-shells
of the conditional average (\ref{eq:ssAvD}). These assumptions
and the equivalence of sites after disorder averaging, lead us to
\begin{eqnarray}
\hspace*{-15mm}
A[s,s^\prime;h,h^\prime;\tilde s]&=&\lim_{N\rightarrow
\infty}\frac{N\!-\!1}{c}\Big\langle\frac{\sum_{\sigmav}\prod_{\tau\mu}\delta
\left[D(\tau,h_{\mu}) \!-\!
D(\tau,h_\mu;\sigmav)\right]}{\sum_{{\sigmav^\prime}}\prod_{\tau\mu}\delta
\left[D(\tau,h_{\mu})\! -\!
D(\tau,h_\mu;{\sigmav^\prime})\right]}
\nonumber \\
\hspace*{-15mm}
&&\times\delta_{{s^\prime},\sigma_1}\delta_{s,\sigma_2}c_{12}\delta
[h^\prime\! - h_1 (\sigmav)]\delta [ h - h_2 (\sigmav)+
2J_{12}\tilde s]\Big\rangle_{\{c_{ij}J_{ij}\}}
\end{eqnarray}
We eliminate the fraction from the above expression via the replica identity
\begin{eqnarray}
\frac{\sum{\sigmav}\Phi
(\sigmav)W(\sigmav)}{\sum_{\sigmav}W(\sigmav)}=\lim_{n\rightarrow
0}\sum_{\sigmav^1}\ldots\sum_{\sigmav^n}\Phi
(\sigmav^1)\prod_{\alpha=1 }^n W(\sigmav^{\alpha})
\end{eqnarray}
which leads to
\begin{eqnarray}
\hspace*{-15mm}
A[s,s^\prime ;h,h^\prime ;\tilde s]
&=&\lim_{N\rightarrow\infty}\lim_{n\rightarrow
0}\frac{N\!-\!1}{c}\Big\langle\sum_{\sigmav^1}\ldots\sum_{\sigmav^n}
\nonumber
\\
\hspace*{-15mm}
&&\times \delta_{{s^\prime},\sigma_1^1}\delta_{s,\sigma_2^1}c_{12}\delta
[\acute h - h_1 (\sigmav^1)]\delta [ h - h_2
(\sigmav^1)+ 2J_{12}\tilde s]\label{eq:A1}~~~~~~~\\
\hspace*{-15mm}
& & \times\prod_{\alpha =1}^n \prod_{\tau\mu}\delta
\Big[D(\tau,h_{\mu})-
\frac{1}{N}\sum_i\delta_{\tau,\sigma_i^{\alpha}}\delta\left[h_{\mu}-
h_i(\sigmav^{\alpha})\right]\Big]\Big\rangle_{\{c_{ij}J_{ij}\}}\nonumber
\end{eqnarray}
We can remove the disorder dependent
local fields $\{h_i(\sigmav^{\alpha})\}$ from inside the delta
functions by  inserting into (\ref{eq:A1}) the following integral representation of unity:
$1=\int\prod_{\alpha i} \mathrm d H_i^{\alpha}\delta
[H_i^{\alpha}- h_i (\sigmav^{\alpha})]$.
Writing the latter delta functions in integral form gives
\begin{eqnarray}
\hspace*{-15mm}
A[s,s^\prime;h, h^\prime;\tilde s]
&=&\lim_{N\rightarrow\infty}\lim_{n\rightarrow
0}\frac{N\!-\!1}{c}\sum_{\sigmav^1}\ldots\sum_{\sigmav^n}\int\prod_{\alpha
 i}\Big[ \mathrm d H_i^{\alpha}\mathrm d
\hat{h}_i^{\alpha}~
\exp[\rmi\hat{h}_i^{\alpha}H_i^{\alpha}]\Big]
\nonumber
\hspace*{-15mm}
\\
& & \times\prod_{\alpha =1}^n \prod_{\tau\mu}\delta \Big[D(\tau,h_{\mu})
-
\frac{1}{N}\sum_i\delta_{\tau,\sigma_i^{\alpha}}\delta[h_{\mu}-
H_i^{\alpha}]\Big]
\label{eq:A2}
\\
\hspace*{-15mm}
& &\hspace*{-15mm}
 \times\delta_{{s^\prime},\sigma_1^1}\delta_{s,\sigma_2^1}\delta
[h^\prime\!\! -\! H_1^1] ~\Big\langle c_{12}\delta [ h\! -\!
H_2^1\!+\! 2J_{12}\tilde s]\rme^{- \rmi\sum_{\alpha
i}\hat{h}_i^{\alpha}h_i
(\sigmav^{\alpha})}
\Big\rangle_{\{c_{ij}J_{ij}\}}\nonumber
\end{eqnarray}
After the average over the disorder is taken (see
\ref{section:average} for details), we then find
\begin{eqnarray}
\hspace*{-15mm}
A[s,s^\prime;h,h^\prime;\tilde s]
&=&\lim_{N\rightarrow\infty}\lim_{n\rightarrow
0}\sum_{\sigmav^1}\ldots\sum_{\sigmav^n}\delta_{{s^\prime},\sigma_1^1}\delta_{s,\sigma_2^1}\int\prod_{\alpha
 i}\Big[\frac{ \mathrm d H_i^{\alpha}\mathrm d
\hat{h}_i^{\alpha}}{2\pi}
\rme^{\rmi\hat{h}_i^{\alpha}H_i^{\alpha}}\Big]\label{eq:A3}
\\
\hspace*{-15mm}
& & \times\delta [h^\prime\! - H_1^1] \prod_{\tau\mu\alpha}\delta
\Big[D(\tau,h_{\mu}) -
\frac{1}{N}\sum_i\delta_{\tau,\sigma_i^{\alpha}}\delta[h_{\mu}\!-
H_i^{\alpha}]\Big]\nonumber
\\
\hspace*{-15mm}
& & \times \rme^{- \rmi\sum_{\alpha  i}\hat{h}_i^{\alpha}\theta}
\int\! \mathrm d J~ P(J)\delta [ h\! -\! H_2^1\!+ \!2J\tilde
s]\rme^{- \rmi J\sum_{\alpha}[\hat{h}_1^{\alpha}
\sigma_2^{\alpha}+ \hat{h}_2^{\alpha}
\sigma_1^{\alpha}]}\nonumber
\\
\hspace*{-15mm}
&&\times\exp \Big[\frac{c}{2N}\sum_{ij}\Big(\int\! \mathrm d J~
P(J)\rme^{- \rmi J\sum_{\alpha}[\hat{h}_i^{\alpha}
\sigma_j^{\alpha}+ \hat{h}_j^{\alpha}
\sigma_i^{\alpha}]}\!-1\Big)+ O(1)\Big]\nonumber
\end{eqnarray}
The $\order(1)$ term in the exponent of the last line is independent of
$\{s,s^\prime,h,h^\prime,\tilde s\}$, and can always be recovered from the normalization
$\sum_{s,s^\prime}\int\!dh dh^\prime A[s,s^\prime;h,h^\prime;\tilde{s}]=1$.
Next we achieve factorization over
sites in (\ref{eq:A3}) upon isolating the density
\begin{eqnarray}
P(\sigmav , \hat{\hv} ;\{\sigmav_i\},\{\hv_i\})=\frac{1}{N}\sum_i
\delta_{\sigmav , \sigmav_i}\delta [\hat{\hv}-\hat{\hv}_i]
\end{eqnarray}
via insertion into (\ref{eq:A3}) of the $\delta$-functional unity
representation
\begin{eqnarray}
1=\int\! \prod_{\sigmav \hat{\hv}}\mathrm d P(\sigmav ,
\hat{\hv})~\delta [P(\sigmav , \hat{\hv})- P(\sigmav ,
\hat{\hv} ;\{\sigmav_i\},\{\hv_i\})]
\end{eqnarray}
which gives, with the short-hands $\bra g(J) \ket_J=\int\!\rmd J~P(J)g(J)$ and $\bx\cdot\by=\sum_\alpha x^\alpha y^\alpha$,
\begin{eqnarray}
\hspace*{-15mm}
 A[s,s^\prime;h,
h^\prime;\tilde s] &=&\lim_{N\rightarrow\infty}\lim_{n\rightarrow
0}\int\!\prod_{\tau\mu\alpha}\Big[\frac{\mathrm
d\hat{D}_{\alpha}(\tau,h_{\mu})}{2\pi/N}\Big]\int\!\prod_{\sigmav
\hat{\hv}}\Big[\frac{\mathrm d \hat{P}(\sigmav , \hat{\hv})\mathrm
d
P(\sigmav , \hat{\hv})}{2\pi/N}\Big]
\nonumber
\\
\hspace*{-15mm}
&&\hspace*{-15mm} \times\exp \Big\{ N\Big[
\rmi\sum_{\tau\mu\alpha}\hat{D}_{\alpha}(\tau,h_{\mu})D(\tau,h_{\mu})+ \rmi
\sum_{\sigmav \hat{\hv}}\hat{P}(\sigmav ,
\hat{\hv})P(\sigmav , \hat{\hv}) + O(\frac{1}{N}) \nonumber
\\
\hspace*{-15mm}
&&
\hspace*{-5mm}
+ \frac{1}{2}c \sum_{\sigmav {\sigmav^\prime}}\int\! \mathrm d
\hat{\hv}\mathrm d {\hat{\hv}^\prime}~P(\sigmav ,
\hat{\hv})P({\sigmav^\prime} ,\hat{\hv}^\prime )\Big\langle
e^{- \rmi J[\hat{\hv}\cdot{\sigmav^\prime}+\hat{\hv}^\prime\!\cdot\sigmav]}\!-1\Big\rangle_J \Big]\Big\}\nonumber
\\
\hspace*{-15mm}
&&
\hspace*{-15mm}
\times\sum_{\sigmav^1}\ldots\sum_{\sigmav^n}\int\prod_{i}\Big[\frac{
\mathrm d \Hv_i\mathrm d
\hat{\hv}_i}{2\pi}\rme^{\rmi\hat{\hv}_i
\cdot[\Hv_i-\thetav]}\Big]\nonumber
\\
\hspace*{-15mm}
& &
\hspace*{-15mm}
 \times \rme^{ - \rmi\sum_{\tau\mu\alpha}
\hat{D}_{\alpha}(\tau,h_{\mu})\sum_i\delta_{\tau,\sigma_i^{\alpha}}\delta\left[h_{\mu}-
H_i^{\alpha}\right]- \rmi\sum_{\sigmav \hat{\hv}}\hat{P}(\sigmav
, \hat{\hv}) \sum_i
\delta_{\sigmav , \sigmav_i}\delta [\hat{\hv}-\hat{\hv}_i]}\nonumber
\\
\hspace*{-15mm}
& &
\hspace*{-15mm} \times\delta_{{s^\prime},\sigma_1^1}\delta_{s,\sigma_2^1}\delta
[h^\prime - H_1^1]
\Big\langle\delta [ h - H_2^1+ 2J\tilde
s]\rme^{- \rmi J[\hat{\hv}_1 \cdot\sigmav_2+\hat{\hv}_2
\cdot\sigmav_{1}]}\Big\rangle_J
\label{eq:A4}
\end{eqnarray}
where $\sigmav=(\sigma_1,\ldots\sigma_n)$,
$\sigmav_i=(\sigma^1_i,\ldots\sigma^n_i)$ and similarly for the
replicated vectors $\hat{\hv}$, etc. We rescale the
conjugate integration variables according to $\hat{P}(\sigmav ,
\hat{\hv})\rightarrow\mathrm d \hat{\hv}~\hat{P}(\sigmav ,
\hat{\hv})$ and $\hat{D}_{\alpha}(\tau,h_{\mu})\rightarrow\Delta
h_{\mu}\hat{D}_{\alpha}(\tau,h_{\mu})$. This
converts the sums over $\hat{\hv}$ and $\mu$ in the exponent of (\ref{eq:A4})
into well-defined integrals when $\mathrm d
\hat{\hv}\rightarrow{\bnull}$ and $\ell\to \infty$.  We write the resulting path integral
measure as $\{\mathrm d P\mathrm d\hat P\mathrm d\hat D\}$. Next we define an effective single-site measure $M$:
\begin{eqnarray}
\Big\langle f[\Hv,\hat{\hv};\sigmav]\Big\rangle_{M}&=&\frac{\sum_{\sigmav}\int\!
\mathrm d \Hv\mathrm d
\hat{\hv}~M[\Hv,\hat{\hv},\sigmav\vert\theta]f[\Hv,\hat{\hv};\sigmav]}{\sum_{\sigmav}\int\!
\mathrm d \Hv\mathrm d
\hat{\hv}~M[\Hv,\hat{\hv},\sigmav\vert\theta]}\label{def:M}
\\
M[\Hv,\hat{\hv},\sigmav\vert\theta]&=&\rme^{\rmi\hat{\hv}
\cdot[\Hv-\thetav]- \rmi\sum_{s\mu\alpha}\Delta
h_{\mu}
\hat{D}_{\alpha}(s,h_{\mu})\delta_{s,\sigma_{\alpha}}\delta[h_{\mu}-
H_{\alpha}] - \rmi\hat{P}(\sigmav , \hat{\hv})}\nonumber
\end{eqnarray}
and the function
\begin{eqnarray}
\hspace*{-15mm}
\Psi[\{P,\hat{P},\hat{D}\}] &=& \rmi \sum_{s\mu\alpha}\Delta
h_{\mu}\hat{D}_{\alpha}(s,h_{\mu})D(s,h_{\mu})+ \rmi
\sum_{\sigmav}\int\! \mathrm d \hat{\hv}~\hat{P}(\sigmav ,
\hat{\hv})P(\sigmav , \hat{\hv})\nonumber
\\
\hspace*{-15mm}
&&+ \log\sum_{\sigmav}\int\! \mathrm d
\Hv\mathrm d \hat{\hv}~M[\Hv,\hat{\hv},\sigmav\vert\theta]
\label{def:Psi1}
\\
\hspace*{-15mm}
&&
+ \frac{1}{2}c \sum_{\sigmav {\sigmav^\prime}}\int\! \mathrm d
\hat{\hv}\mathrm d \hat{\hv}^\prime~P(\sigmav ,
\hat{\hv})P({\sigmav^\prime} , \acute{\hat{\hv}} )\Big\langle
\rme^{- \rmi J[\hat{\hv} \cdot{\sigmav^\prime}+\hat{\hv}^\prime
\!\cdot\sigmav]}-1\Big\rangle_J\nonumber
\end{eqnarray}
Using these definitions and changing the order of the limits
$N\rightarrow\infty$ and $n\rightarrow 0$ allows us to write
(\ref{eq:A4}) in the form
\begin{eqnarray}
\hspace*{-15mm}
A[s,s^\prime;h,
h^\prime;\tilde s] &=&\lim_{n\rightarrow
0}\lim_{N\rightarrow\infty}\int\!\{\mathrm d P\mathrm d\hat P\mathrm d\hat
D\}~\rme^{N\Psi[\{P,\hat{P},\hat{D}\}]+ O(1)}\label{eq:A5}\\
\hspace*{-15mm}
&&\times\Big\langle\delta_{{s^\prime}\!,\sigma_1}\delta_{s,\sigma^\prime_1}\delta
[h^\prime\! - H_1]
\delta [ h - {H}^\prime_1\!+2J\tilde
s]\rme^{- \rmi J[\hat{\hv} \cdot{\sigmav^\prime}+\hat{\hv}^\prime\!\cdot
\sigmav}\Big\rangle_{J,M,M^\prime}\nonumber
\end{eqnarray}
Finally, with the help of the normalization identity
$\sum_{ss^\prime}\int\!\mathrm d h \mathrm d h^\prime
A[s,s^\prime;h,h^\prime;\tilde s]=1$, we compute (\ref{eq:A5}) by steepest descent:
\begin{eqnarray}
\hspace*{-15mm}
A[s,s^\prime;h,h^\prime;\tilde s]&=&\lim_{n\rightarrow 0}\frac{\big\langle\delta_{{s^\prime}\!,\sigma_1}\delta_{s,\sigma^\prime_1}\delta
[h^\prime\!\! -\! H_1]\delta [ h\! -\! H^\prime_1\!+\! 2J\tilde
s]\rme^{- \rmi J[\hat{\hv} \cdot{\sigmav^\prime}+\hat{\hv}^\prime
\cdot\sigmav]}\big\rangle_{J,M,\acute M }}{\big\langle \rme^{- \rmi
J[\hat{\hv} \cdot {\sigmav^\prime}+\hat{\hv}^\prime\!
\cdot\sigmav]}\big\rangle_{J,M,\acute M}}
\nonumber\\[-2mm]
\hspace*{-15mm}&&
\label{eq:A}
\end{eqnarray}
where $\{P,\hat{P},\hat{D}\}$ are determined by extremization of $\Psi$.
The functional variation of $\Psi$ with respect to $P(\sigmav ,
\hat{\hv}), \hat{P}(\sigmav , \hat{\hv})$ and
$\hat{D}_{\alpha}(s,h_{\mu})$ leads to the stationarity conditions
\begin{eqnarray}
D(s,h) &=& \langle
\delta_{s,\sigma_{\alpha}}\delta
[h-H_{\alpha}]\rangle_{M}
\label{eq:D}
\\
P(\sigmav , \hat{\hv}) &=&\langle
\delta_{\sigmav,{\sigmav^\prime}}\delta
[\hat{\hv}-\hat{\hv}^\prime]\rangle_{\acute M}
\label{eq:P}
\\
\hat{P}(\sigmav , \hat{\hv}) &=&\rmi c
\sum_{{\sigmav^\prime}}\int\!\mathrm d
\hat{\hv}^\prime~ P({\sigmav^\prime} ,\hat{\hv}^\prime)\langle
e^{-\rmi J[\hat{\hv} \cdot{\sigmav^\prime}+\hat{\hv}^\prime\!\cdot
\sigmav]}-1\rangle_J
\label{eq:Pconjugate}
\end{eqnarray}
The conjugate order parameters $\hat{D}_{\alpha}(s,h)$ and $\hat{P}(\sigmav , \hat{\hv})$ are seen to play the role of Lagrange multipliers, ensuring normalization of $D(s,h)$ and $P(\sigmav , \hat{\hv})$. The physical meaning of the density $P(\sigmav , \hat{\hv})$
is not yet clear, due to the presence of the vector $\hat{\hv}$.

We use equation (\ref{eq:Pconjugate}) to eliminate the conjugate
order parameters $\hat{P}(\sigmav , \hat{\hv})$ from the measure $M$.
 We assume that $\hat{D}(s,h)$ is sufficiently smooth in $h$, such that $\sum_{\mu}\Delta
h_{\mu}\hat{D}_{\alpha}(s,h_{\mu})f(h_{\mu})\rightarrow\int\!\mathrm d
H~\hat{D}_{\alpha}(s,H)f(H)$ for $\ell\rightarrow\infty$. This leads to
\begin{eqnarray}
M[\Hv,\hat{\hv},\sigmav\vert\theta]&=&\exp\Big\{\rmi\hat{\hv}
\cdot[\Hv-\thetav]- \rmi\sum_{\alpha}
\hat{D}_{\alpha}(\sigma_{\alpha},H_{\alpha})
\nonumber
\\
&&+ c
\sum_{{\sigmav^\prime}}\int\!\mathrm d
\hat{\hv}^\prime~P({\sigmav^\prime} ,\hat{\hv}^\prime )\big\langle
e^{- \rmi J[\hat{\hv} \cdot{\sigmav^\prime}+\hat{\hv}^\prime\!\cdot
\sigmav]}\!- 1\big\rangle_J \Big\}\label{eq:M1}~~~~~~
\end{eqnarray}
in the definition of the measure $M$ (\ref{def:M}).
%and
%
%\begin{eqnarray}
%\Psi[\{P,\hat{D}\}] &=& i
%\sum_{s,\alpha}\int\mathrm d
%H\hat{D}_{\alpha}(s,H)D(s,H)+\log\sum_{\sigmav}\int
%\mathrm d \Hv\mathrm d \hat{\hv}M[\Hv,\hat{\hv},\sigmav\vert\theta]\label{eq:Psi2}~~~~~~~~\\
%&&-\frac{1}{2}c \sum_{\sigmav,{\sigmav^\prime}}\int \mathrm d
%\hat{\hv}\mathrm d \acute{\hat{\hv}}P(\sigmav ,
%\hat{\hv})P({\sigmav^\prime} , \acute{\hat{\hv}})\left\langle
%e^{- i J\hat{\hv} .{\sigmav^\prime}- i J\acute{\hat{\hv}}
%.\sigmav}- 1\right\rangle_J\nonumber
%\end{eqnarray}
%
The replica method requires finally that we take the $n\rightarrow 0$
limit in equations (\ref{eq:A}-\ref{eq:P}). To do this we need
to make appropriate \emph{ans\"{a}tze} for the density $P(\sigmav , \hat{\hv})$
and for the conjugate order parameters $\hat{D}_{\alpha}(s,H)$.

\subsection{Replica symmetry}\label{section:RS}
We evaluate (\ref{eq:A})-(\ref{eq:P}) upon assuming ergodicity, which translates mathematically into the so-called replica-symmetry (RS)
ansatz. Firstly, the order
parameters $\hat{D}_{\alpha}(s,H)$ depend only on a single replica index and are expected to be imaginary, so we put
% For the conjugate
%parameter $\hat{D}_{\alpha}(s,H)$ the RS simply implies
%loosing its dependence on the replica index $\alpha$. Assuming that
%$\hat{D}_{\alpha}(s,H)$ is purely imaginary this gives
%
\begin{eqnarray}
\hat{D}_{\alpha}(s,H)&=&\rmi\log d(s,H)\label{def:Drs}
\end{eqnarray}
Second, the density $P(\sigmav , \hat{\hv})$ depends on a discrete and continuous vector in replica space. The
RS ansatz demands its invariance
under any joint permutation of their indices, which implies
\cite{SoftSpins} that it must be of the general form
%within RS the density $P(\sigmav , \hat{\hv})$ takes the form
%
\begin{eqnarray}
P_{RS}(\sigmav, \hat{\hv})&=&\int\! \left\lbrace \mathrm d
P\right\rbrace~ W[\{P\}]\prod_{\alpha=1}^n
P(\sigma_{\alpha},\hat{h}_{\alpha})\label{def:Prs}
\end{eqnarray}
 where $W[\{P\}]$ is a normalized functional distribution, i.e. $\int\! \left\lbrace \mathrm d
P\right\rbrace W[\{P\}]=1$.
 The RS ansatz (\ref{def:Drs},\ref{def:Prs}), via
 its implications for the effective measure (\ref{def:M}), will enable us to take the replica limit $n\to 0$ in equations (\ref{eq:A}-\ref{eq:P}). We insert (\ref{def:Drs},\ref{def:Prs}) into (\ref{eq:M1}) and subsequently
  expand the exponential
  function containing $P_{RS}(\sigmav, \hat{\hv})$, leading to
\begin{eqnarray}
\hspace*{-15mm}
M_{RS}[\Hv,\hat{\hv},\sigmav\vert\theta] &=& \sum_{k\geq
0}\frac{c^k}{k!}e^{- c}\int\!\prod_{\ell=1}^k\Big\lbrace \mathrm d
J_\ell P(J_\ell)
\lbrace \mathrm d P_\ell\rbrace W[\{P_l\}]\Big\rbrace
\label{eq:Mrs}
\hspace*{-15mm}
\\
&&\hspace*{-21mm} \times \prod_{\alpha=1}^n\Big\{
d(\sigma_{\alpha},H_{\alpha})e^{\rmi\hat{h}_{\alpha}[
H_{\alpha}- \theta]}
 \prod_{\ell=1}^k\Big[
\sum_{\sigma^{\alpha}_{\ell}}\int\!\mathrm d \hat{h}^{\alpha}_{\ell}
P_\ell(\sigma^{\alpha}_{\ell},\hat{h}^{\alpha}_{\ell})e^{- \rmi
J_\ell[\hat{h}_{\alpha}\sigma^{\alpha}_\ell+\hat{h}^{\alpha}_{\ell}
\sigma_{\alpha}]}\Big] \Big\}\nonumber
\end{eqnarray}
We write averages with respect to the RS measure (\ref{eq:Mrs}) as
\begin{eqnarray}
\hspace*{-5mm}
\langle f[\Hv,\hat{\hv};\sigmav]\rangle_{M_{RS}}&=&\frac{1}{M_{RS}^{n}}\sum_{\sigmav}\int\!
\mathrm d \Hv\mathrm d
\hat{\hv}~M_{RS}[\Hv,\hat{\hv},\sigmav\vert\theta]f[\Hv,\hat{\hv};\sigmav]
\end{eqnarray}
where we defined the normalization constant $M_{RS}^{n}=\sum_{\sigmav}\int\! \mathrm d
\Hv\mathrm d \hat{\hv}~M_{RS}[\Hv,\hat{\hv},\sigmav\vert\theta]$. Clearly $\lim_{n\rightarrow 0}M_{RS}^{n}=1$.
We use the above results to solve equation (\ref{eq:P}) for the functional distribution $W[\{P\}]$, upon substituting
 the various RS expressions:
\begin{eqnarray}
\hspace*{-15mm}&& \hspace*{-25mm}
M_{RS}^{n}\int\! \left\lbrace \mathrm d P\right\rbrace
W[\{P\}]\prod_{\alpha=1}^n
P(\sigma_{\alpha},\hat{h}_{\alpha})
=\sum_{k\geq 0}\frac{c^k}{k!}e^{-
c}\int\!\prod_{\ell=1}^k\Big\{ \mathrm d J_\ell P(J_\ell)
\left\lbrace \mathrm d P_\ell\right\rbrace W[\{P_\ell\}]\Big\}
 \nonumber\\
 \hspace*{-15mm}
&&
\hspace*{-15mm}
\times \prod_{\alpha=1}^n\int\! \mathrm d
H_{\alpha}d(\sigma_{\alpha},H_{\alpha})\rme^{\rmi\hat{h}_{\alpha}[
H_{\alpha}-\theta]}
\prod_{\ell=1}^k\Big[
\sum_{\sigma^{\alpha}_{\ell}}\int\!\mathrm d \hat{h}^{\alpha}_{\ell}
P_\ell(\sigma^{\alpha}_{\ell},\hat{h}^{\alpha}_{\ell})\rme^{- \rmi
J_\ell[\hat{h}_{\alpha}\sigma^{\alpha}_\ell+\hat{h}^{\alpha}_{\ell} \sigma_{\alpha}]}\Big]
 \nonumber\\
&=&
\int\!\left\lbrace \mathrm d P\right\rbrace \prod_{\alpha=1}^n
P(\sigma_{\alpha},\hat{h}_{\alpha})
 \sum_{k\geq
0}\frac{c^k}{k!}e^{- c}\int\!\prod_{\ell=1}^k\Big\{ \mathrm d
J_\ell P(J_\ell)
\left\lbrace \mathrm d P_\ell\right\rbrace W[\{P_\ell\}]\Big\}
\nonumber\\
&&\times Z^n [\{P_1,\ldots,P_k\}]\nonumber\\
&&\hspace*{-20mm}
\times\prod_{\sigma\hat h}\delta\Big[
 P(\sigma,\hat h)
-\frac{\int\! \mathrm d H d(\sigma,H)\rme^{\rmi\hat{h}[
H-\theta]}\prod_{\ell=1}^k\{
\sum_{\sigma_{\ell}}\int\!\mathrm d \hat{h}_{\ell}
P_\ell(\sigma_{\ell},\hat{h}_{\ell})\rme^{- \rmi J_\ell[\hat{h}\sigma_{\ell}+\hat{h}_{\ell}
\sigma]}\}}{Z[\{P_1,\ldots,P_k\}]}\Big]\nonumber
\end{eqnarray}
where
\begin{eqnarray}
Z[\{P_1,\ldots,P_k\}]&=&\sum_{\sigma}\int\! \mathrm d H\mathrm d \hat
h~ d(\sigma,H)\rme^{\rmi\hat{h}[H-\theta]}\nonumber
\\
&&\times \prod_{\ell=1}^k\Big\{
\sum_{\sigma_{\ell}}\int\!\mathrm d \hat{h}_{\ell}
P_\ell(\sigma_{\ell},\hat{h}_{\ell})\rme^{- \rmi J_l[\hat{h}\sigma_{\ell}+\hat{h}_{\ell} \sigma]}\Big\}\label{def:Z}~~~~~~~~~~
\\
\hspace*{-15mm}
&&\hspace*{-15mm}= 2\pi\sum_{\sigma}\!\prod_{\ell=1}^k\Big\{
\sum_{\sigma_{\ell}}\int\!\mathrm d \hat{h}_{\ell}
P_\ell(\sigma_{\ell},\hat{h}_{\ell})\rme^{- \rmi J_\ell\hat{h}_{\ell}
\sigma}\Big\} d\big(\sigma,\!\sum_{\ell=1}^k
J_\ell\sigma_{\ell}\!+\!\theta\big) \nonumber
\end{eqnarray}
In the limit $n\rightarrow 0$ both the normalization term $Z^n
[\{P_1,\ldots,P_k\}]$ and the constant $M_{RS}^{n}$ reduce
to unity, and we find an equation for the functional distribution $W[\{P\}]$:
\begin{eqnarray}
\hspace*{-20mm}
W[\{P\}]&=& \sum_{k\geq
0}\frac{c^k}{k!}e^{- c}\int\!\prod_{\ell=1}^k\Big\{ \mathrm d
J_\ell P(J_\ell)
\left\lbrace \mathrm d P_\ell\right\rbrace W[\{P_\ell\}]\Big\}
\label{eq:W}\\
\hspace*{-20mm}
&&
\hspace*{-15mm}
\times
\prod_{\sigma\hat h}\delta\Big[
 P(\sigma,\hat h)
-\frac{\int\! \mathrm d H d(\sigma,H)\rme^{\rmi\hat{h}[
H-\theta]}\prod_{\ell=1}^k\{
\sum_{\sigma_{\ell}}\int\!\mathrm d \hat{h}_{\ell}
P_\ell(\sigma_{\ell},\hat{h}_{\ell})\rme^{- \rmi J_\ell[\hat{h}\sigma_{\ell}+\hat{h}_{\ell}
\sigma]}\}}{Z[\{P_1,\ldots,P_k\}]}\Big]\nonumber
\end{eqnarray}
In a similar fashion (see \ref{section:RSeq} for details) we can compute the probability distributions $D(s,h)$ and $A[s,s^\prime;h,h^\prime;\tilde s]$ in RS ansatz. To compactify our formulae we define the Fourier transforms $\hat{P}(\sigma|x)=\int\! \mathrm d \hat h
~P(\sigma, \hat h)\rme^{- \rmi\hat h x}$, in terms of which we find
\begin{eqnarray}
D(s,h) &=& ~d(s,h)
\sum_{k\geq
0}\frac{c^k}{k!}e^{- c}\int\!\prod_{\ell=1}^k\Big\{ \mathrm d
J_\ell P(J_\ell)
\left\lbrace \mathrm d P_\ell\right\rbrace W[\{P_\ell\}]\Big\}
\nonumber
 \\
&&
\times \frac{\prod_{\ell=1}^k\big\{ \sum_{\sigma_{\ell}}\hat{P}_\ell(\sigma_{\ell}|J_{\ell}s)\big\} \delta
[h\!-\!\sum_{\ell=1}^k J_\ell\sigma_{\ell}\!-\! \theta]}{Z[\{P_1,\ldots,P_k\}]}
 \label{eq:Drs1}
\end{eqnarray}
\begin{eqnarray}
\hspace*{-20mm}
A[s,s^\prime;h,h^\prime;\tilde s]&=&\sum_{k\geq
0}\frac{c^k}{k!}e^{- c}\int\!\prod_{\ell=1}^k\Big\{ \mathrm d
J_\ell P(J_\ell)
\left\lbrace \mathrm d P_\ell\right\rbrace W[\{P_\ell\}]\Big\}
\nonumber
\\
\hspace*{-20mm}
&&
\times
\sum_{m\geq 0}\frac{c^m}{m!}e^{-
c}\int\!\prod_{r=1}^m\Big\{ \mathrm d J^\prime_r  P(J^\prime_r)
\left\lbrace \mathrm d Q_r\right\rbrace W[\{Q_r\}]\Big\}\nonumber
\\
\hspace*{-20mm}
&&
\times
\left\langle
\prod_{\ell=1}^k\!\left\lbrace
\sum_{\sigma_{\ell}}\!
\hat{P}_\ell(\sigma_{\ell}|J_{\ell}s^\prime)\right\rbrace
\delta [h^\prime \!-\! \sum_{\ell=1}^k
J_\ell\sigma_{\ell}\!-\! \theta\!-\! Js]
d(s^\prime,h^\prime)
\right.\nonumber
\\
\hspace*{-20mm}&&
\left.\times
\prod_{r=1}^m\!\left\lbrace \sum_{\sigma_{r}}\! \hat{Q}_r(\sigma_{r}|J^\prime_r s)\right\rbrace
 \delta [ h\!-\!\sum_{r=1}^m\!
J^\prime_r\sigma_{r}\!-\! \theta\!-\! Js^\prime\!+\! 2J\tilde s]
d(s,h\!+\!2J\tilde{s})
\right.\nonumber
\\
\hspace*{-20mm}
&&\left.
\times\left[
\sum_{\sigma\sigma^\prime} \prod_{\ell=1}^k\!\left\lbrace
\sum_{\sigma_{\ell}}\!
\hat{P}_\ell(\sigma_{\ell}|J_\ell\sigma)\right\rbrace  d\big(\sigma,\sum_{\ell=1}^k
J_\ell\sigma_{\ell}\!+\! \theta\!+\! J \sigma^\prime\big)
\right.\right.
\nonumber
\\
\hspace*{-20mm}
&&\left.\left.\hspace*{5mm}\times
\prod_{r=1}^m\!\left\lbrace \sum_{\sigma_{r}}\! \hat{Q}_r(\sigma_{r}|J^\prime_r\sigma^\prime)\right\rbrace  d\big(\sigma^\prime\!,\sum_{r=1}^m
J^\prime_r\sigma_{r}\!+\! \theta\!+\! J \sigma\big) \right]^{-1}
\right\rangle_{\!J}
\label{eq:Ars1}
\end{eqnarray}
 Our theory requires solution of the saddle-point equations (\ref{eq:W}-\ref{eq:Drs1}) for the functional distribution $W[\{P\}]$ and the function $d(s,h)$. These are functional relations, and is generally not possible to solve them analytically. Furthermore, the imaginary arguments in (\ref{eq:W}) induce further complications in numerical solution.
To simplify matters we assume that for $x\in{\rm I\!R}$ the Fourier transforms $\hat{P}(\sigma|x)$ are real-valued, and we define a corresponding functional distribution
\begin{eqnarray} \tilde W[\{\hat P\}]&=&\int\left\lbrace \mathrm d
P\right\rbrace W[\{P\}]\prod_{\sigma x}
\delta\Big[ \hat P(\sigma \vert x)- \int\! \mathrm d \hat h
~P(\sigma, \hat h)\rme^{- \rmi\hat h x}\Big]\label{def:Wft}
\end{eqnarray}
$\tilde{W}$
 is normalized by construction, but the $\hat P(\sigma \vert x)$ need not be. We transform our problem into the language of $\tilde{W}$ by inserting (\ref{eq:W}) into (\ref{def:Wft}) and integrating over $\{P\}$:
\begin{eqnarray}
\hspace*{-15mm}
\tilde W[\{\hat P\}]&=& \sum_{k\geq 0}\frac{c^k}{k!}e^{-
c}\int\prod_{\ell=1}^k\left\lbrace \mathrm d J_\ell P(J_\ell)
\{\mathrm d \hat P_\ell\} \tilde W[\{\hat P_\ell\}]\right\rbrace\label{eq:Wft}
\\
\hspace*{-15mm}
&&\times\prod_{\sigma x}\delta\left[ \hat P(\sigma \vert x)-
\frac{\prod_{\ell=1}^k\left\lbrace \sum_{\sigma_{\ell}} \hat
P_\ell(\sigma_{\ell}\vert J_\ell\sigma)\right\rbrace d(\sigma,\sum_{\ell=1}^k
J_\ell\sigma_{\ell}\!+\! \theta\!+\! x)}{Z[\{\hat P_1,\ldots,\hat
P_k\}]}\right]\nonumber
\end{eqnarray}
Our previous results
(\ref{eq:Drs1}-\ref{eq:Ars1}) now take the form
\begin{eqnarray}
D(s,h) &=& d(s,h) \sum_{k\geq 0}\frac{c^k}{k!}e^{-
c}\int\prod_{\ell=1}^k\Big\{ \mathrm d J_\ell P(J_\ell)
\{\mathrm d \hat P_\ell\} \tilde W[\{\hat P_\ell\}]\Big\}\label{eq:Drs2}\\
&&\times \frac{\prod_{\ell=1}^k\left\lbrace
\sum_{\sigma_{\ell}} \hat P_\ell(\sigma_{\ell}\vert J_\ell s)\right\rbrace
\delta [h\!-\! \sum_{\ell=1}^k
J_\ell\sigma_{\ell}\!-\! \theta]}{Z[\{\hat P_1,\ldots,\hat
P_k\}]}\nonumber
\end{eqnarray}%
and
\begin{eqnarray}
\hspace*{-20mm}
A[s,s^\prime;h,h^\prime;\tilde s]&=&\sum_{k\geq
0}\frac{c^k}{k!}e^{- c}\int\!\prod_{\ell=1}^k\Big\{ \mathrm d
J_\ell P(J_\ell)
\left\lbrace \mathrm d \hat{P}_\ell\right\rbrace \tilde{W}[\{\hat{P}_\ell\}]\Big\}
\nonumber
\\
\hspace*{-20mm}
&&
\times
\sum_{m\geq 0}\frac{c^m}{m!}e^{-
c}\int\!\prod_{r=1}^m\Big\{ \mathrm d J^\prime_r d P(J^\prime_r)
\left\lbrace \mathrm d \hat{Q}_r\right\rbrace \tilde{W}[\{\hat{Q}_r\}]\Big\}\nonumber
\\
\hspace*{-20mm}
&&
\times
\left\langle
\prod_{\ell=1}^k\!\left\lbrace
\sum_{\sigma_{\ell}}\!
\hat{P}_\ell(\sigma_{\ell}|J_{\ell}s^\prime)\right\rbrace
\delta [h^\prime \!-\! \sum_{\ell=1}^k
J_\ell\sigma_{\ell}\!-\! \theta\!-\! Js]
d(s^\prime,h^\prime)
\right.\nonumber
\\
\hspace*{-20mm}&&
\left.\times
\prod_{r=1}^m\!\left\lbrace \sum_{\sigma_{r}}\! \hat{Q}_r(\sigma_{r}|J^\prime_r s)\right\rbrace
 \delta [ h\!-\!\sum_{r=1}^m\!
J^\prime_r\sigma_{r}\!-\! \theta\!-\! Js^\prime\!+\! 2J\tilde s]
d(s,h\!+\!2J\tilde{s})
\right.\nonumber
\\
\hspace*{-20mm}
&&\left.
\times\left[
\sum_{\sigma\sigma^\prime} \prod_{\ell=1}^k\!\left\lbrace
\sum_{\sigma_{\ell}}\!
\hat{P}_\ell(\sigma_{\ell}|J_\ell\sigma)\right\rbrace  d\big(\sigma,\sum_{\ell=1}^k
J_\ell\sigma_{\ell}\!+\! \theta\!+\! J \sigma^\prime\big)
\right.\right.
\nonumber
\\
\hspace*{-20mm}
&&\left.\left.\hspace*{5mm}\times
\prod_{r=1}^m\!\left\lbrace \sum_{\sigma_{r}}\! \hat{Q}_r(\sigma_{r}|J^\prime_r\sigma^\prime)\right\rbrace  d\big(\sigma^\prime\!,\sum_{r=1}^m
J^\prime_r\sigma_{r}\!+\! \theta\!+\! J \sigma\big) \right]^{-1}
\right\rangle_{\!J}
\label{eq:Ars2}
\end{eqnarray}
Equations (\ref{eq:Wft}-\ref{eq:Ars2}) are the final analytic results within
the replica symmetry theory. They complement and close the diffusion equation
(\ref{eq:diffusion}). We may now proceed to the solution of (\ref{eq:diffusion}) by iterating the following recipe from time $t=0$ onwards:
at any time point $t$ we use
the instantaneous distribution $D_t(s,h)$ to solve equations (\ref{eq:Wft}-\ref{eq:Drs2}) numerically
for $\tilde W[\{\hat P\}]$
and $d(s,h)$ via a population dynamics algorithm
\cite{BetheSG}, the result of which is then used to compute the kernel (\ref{eq:Ars2}), and to iterate
(\ref{eq:diffusion}) over the next infinitesimal time step $t\to t+dt$.
\section{Application of the theory to the vertex cover problem}\label{section:VC}
In this section we show how the theory developed in the previous sections allows us to study analytically the Monte Carlo dynamics (extended with
appropriate stochastic cooling schedules of the simulated annealing type) of the so-called minimum vertex cover optimization problem.
%In this section we apply the theory developed in sections \ref{section:Model}-\ref{section:RS} to describe the dynamics of minimum vertex cover optimization problem.
%
\subsection{The minimal vertex cover problem}
We start with the definitions. Let $G=(V,E)$ be a graph defined by a set of $N$ vertices $V=\{1,\ldots,N\}$ and a set of undirected edges  $E=\{(i,j)\}$, where $i,j\in V$ and there is no distinction between
$(i,j)$ and $(j,i)$. A vertex cover (VC) of a graph $G$ is a subset $V_{VC}\subseteq V$ of vertices, such that for all edges $(i,j)\in E$
either $i\in V_{VC}$ or $j\in V_{VC}$ or both. The vertices in $V_{VC}$ are called covered, and those in $V\setminus V_{VC}$ uncovered.
Similarly, an edge $(i,j)$ is said to be covered if at least one of the vertices $\{i,j\}$ is in $V_{VC}$.
The \emph{minimum vertex cover} problem is the following optimization problem: find a vertex cover set $V_{VC}$ of {\em minimal} cardinality, for a given graph $G$, and compute the fraction $x_c(G)=\vert V_{VC}\vert/N$. The corresponding \emph{decision problem}, to find whether for a given graph $G$ a VC  of fixed cardinality $x=\vert V_{VC}\vert/N$ exists, is known to belong to the class of NP-complete problems \cite{Garey}; i.e. it is conjectured
that no algorithm of polynomial time complexity in $N$ (or in the number of edges $M$) exists to solve it. All algorithms known to date indeed have exponential time complexity.
The introduction of graph ensembles allows to study typical instances of the minimal VC problem, and quantify average properties. For instance, it was found that for large Poissonnian random graphs (\ref{eq:P(c)}) the fraction of covered vertices in a minimal vertex cover $x_c(G)$ depends on the average connectivity $c$ only, i.e. $x_c(G)=x_c(c)$. Rigorous lower and upper bounds for $x_c(c)$ were derived in \cite{Gazmuri}:
\begin{eqnarray}
x_l(c)<x_c(c)<1-\ln(c)/c\label{eq:bounds}
\end{eqnarray}
where $x_l(c)$ is a solution of
\begin{eqnarray}
x \ln x+ (1- x)\ln(1-x)-\frac{1}{2}c(1- x)^2=0\label{eq:lbound}
\end{eqnarray}
The lower bound coincides with the annealed bound calculated within statistical mechanics, see e.g. \cite{VCising2}.
The asymptotic form of $x_c(c)$ for large $c$ was given in \cite{Frieze}:
\begin{eqnarray}
x_c(c)=1-2[1+\ln(c)-\ln(\ln (c)) -\ln(2)]/c+o(c^{-1})
\label{eq:asymptotic}
\end{eqnarray}
Since in VC problems a vertex is either covered or uncovered, one can map the VC problem in to an Ising model \cite{VCising1}\footnote{
An alternative representation involves mapping the VC problem into a hard-sphere lattice gas \cite{VCgas}.}:
for any subset $U\subseteq V$ we define $\sigma_i=1$ if $i\in U$,  and $\sigma_i=-1$ if $i\notin U$.  We define a corresponding
Hamiltonian for the state $\bsigma=(\sigma_1,\ldots,\sigma_N)$ which simply counts the number of uncovered edges:
\begin{eqnarray}
H(\sigmav)=\sum_{i<j}c_{ij}\delta_{\sigma_i,-1}\delta_{\sigma_j,-1}\label{def:VCenergy1}
\end{eqnarray}
Solving the VC problem for a given relative cardinality $x$ then reduces to
minimizing $H(\bsigma)$ under the constraint $\sum_i\delta_{\sigma_i,1}=xN$, which in terms of the Ising spins implies
\begin{eqnarray}
\frac{1}{N}\sum_i\sigma_i=2x-1\label{def:constraint}
\end{eqnarray}
The study of VC has thereby been connected to the study of the ground states of an Ising spin system,
within equilibrium statistical mechanics. This enabled the computation of the relative size of  minimal VCs
for typical large graph instances of FC ensembles, by averaging the free energy of the spin system over all graphs in the ensemble. Within the RS ansatz this resulted for Poisonnian graphs in
\cite{VCising1,VCgas}:
\begin{eqnarray}
x_c(c)=1-\frac{2W(c)+W^2(c)}{2c}\label{eq:treshold}
\end{eqnarray}
Here $W(c)$ is the Lambert $W$-function \cite{Lambert}, defined as the real solution of $c=We^{W}$. This result implies that almost all graphs are coverable with $xN$ vertices for $x>x_c(c)$ and not coverable for $x<x_c(c)$. It was proved to be exact for $c\leq\exp(1)$ \cite{VCrig}, but for  $c>\exp(1)$ equation (\ref{eq:treshold}) underestimates the empirical values of $x_c(c)$ obtained by numeric simulation \cite{VCising1}, and for $c\geq20.7$ it even violates the lower bound  (\ref{eq:lbound}). The explanation was found to be that at $c=\exp(1)$ the
assumed RS breaks down \cite{VCising1,VCgas}, and replica symmetry breaking (RSB) occurs. The one-step RSB solution was computed in \cite{Zhou} via the cavity method; its agreement with the numeric results \cite{VCising1} improves upon the RS calculation (\ref{eq:treshold}) and approaches (\ref{eq:asymptotic}) correctly for large $c$, yet the one-step RSB solution is still incorrect for $c>\exp(1)$ \cite{Zhou}. A more recent result obtained in \cite{LongRange} is in good agreement with both the numerical simulations \cite{VCising1} for $c\leq10.0$, and with the asymptotic form (\ref{eq:asymptotic}) for large $c$.
 The few and limited
 analytic studies of the {\em dynamics} of algorithmic solutions of the VC problem were carried out only for a simple backtracking algorithm \cite{VCdyn1}, and for more complex heuristic \cite{VCdyn2} algorithms. In this paper, in contrast, we consider a more physical dynamics, inspired by the connection with a ground state search in Ising spin systems.

\subsection{Dynamic replica analysis of the vertex cover problem}
In this section we analyze a Monte Carlo dynamics for the minimum VC problem of the type (\ref{eq:master}),
where we allow the temperature $T$ to vary over time such that $\lim_{t\to\infty}T(t)=0$, but sufficiently slowly so that in (\ref{eq:master}) we may simply substitute
$\beta \to\beta(t)$.
 We map the VC problem into the Ising model (\ref{def:VCenergy1}), and impose the constraint (\ref{def:constraint}) in a `soft' way,
 by adding an extra term to the Hamiltonian (\ref{def:VCenergy1})
\begin{eqnarray}
\tilde H(\sigmav)=\sum_{i<j}c_{ij}\delta_{\sigma_i,-1}\delta_{\sigma_j,-1}-\lambda\sum_i\delta_{\sigma_i,-1}\label{def:VCenergy2}
\end{eqnarray}
(where $\lambda>0$),
 which ensures that among the states $\bsigma$ that minimize (\ref{def:VCenergy2}),  those  with the smallest sum $\sum_i\delta_{\sigma_i,1}$ (i.e. minimal cover) are preferred. The Glauber dynamics associated with the Hamiltonian (\ref{def:VCenergy2}) is indeed
 of the type (\ref{eq:algorithm}), with the local field
\begin{eqnarray}
h_i(\sigmav)=J\sum_{j\neq i} c_{ij}\delta_{\sigma_j,- 1}+ \theta\label{eq:VCfield}
\end{eqnarray}
where $J=\frac{1}{2}$ and $\theta=-\frac{1}{2}\lambda$.
The fields could also have been written in the more standard form $h_i(\bsigma)=\sum_j J_{ij}\sigma_j+\theta_i$, but this would have required
site dependent random $\theta_i$ which involve the connectivity variables $c_{ij}$.
The consequences for our theory of changing the local fields from the conventional form (\ref{eq:field}) to (\ref{eq:VCfield}) are minor.
The diffusion equation (\ref{eq:diffusion}) remains unchanged; the only difference is in definition of the distribution (\ref{eq:jsfield}), which now
becomes
\begin{eqnarray}
D(s,h;\sigmav)&=&\frac{1}{N}\sum_i\delta_{s,\sigma_i}\delta[h- J\sum_{j\neq i} c_{ij}\delta_{\sigma_j,- 1}- \theta]\label{eq:VCjsfield}
\end{eqnarray}
and the kernel (\ref{def:A}), which changes to
\begin{eqnarray}
\hspace*{-15mm}
\tilde{A}[s,s^\prime;h,h^\prime;\tilde s]=\Big\langle\frac{1}{c
N}\sum_{i
j}\delta_{s^\prime\!,\sigma_i}\delta_{s,\sigma_j}c_{ij}~\delta
[h^\prime\! -\! h_i (\sigmav)]\delta [ h\! -\! h_j (\sigmav)\!-\! J\tilde
s]\Big\rangle_{D;t}\label{def:Avc}
\end{eqnarray}
where $\tilde s \in \lbrace 0,s^\prime\rbrace$ and $h_i (\sigmav)$ is given by (\ref{eq:VCfield}).
 Next we compute the consequences of defining (\ref{eq:VCfield}) within the  replica calculations. This involves only minor alterations
 of the steps taken in section \ref{section:RA}, and we
readily obtain the new expression that replaces our previous (\ref{eq:A}) (where in VC there is of course no longer a need to average over $J$):
\begin{eqnarray}
\hspace*{-15mm}
A[s,s^\prime;h,h^\prime;\tilde s]&=&
\label{eq:Avc}
\\
\hspace*{-15mm}&&
\hspace*{-15mm}\lim_{n\rightarrow 0}
\frac{\big\langle
\delta_{s^\prime\!,\sigma_1}\delta_{s,\sigma^\prime_1}\delta
[h^\prime\!\! -\! H_1]\delta [ h\! -\! H^\prime_1\!\!-\! J\tilde s]\rme^{- \rmi J\sum_{\alpha}[\hat{h}_{\alpha}
\delta_{\sigma^\prime_{\alpha},- 1}+ \hat{h}^\prime_{\alpha}
\delta_{\sigma_{\alpha},- 1}]}
\big\rangle_{M,\acute M }}{\big\langle \rme^{- \rmi J\sum_{\alpha}[\hat{h}_{\alpha}
\delta_{\sigma^\prime_{\alpha},- 1}+ \hat{h}^\prime_{\alpha}
\delta_{\sigma_{\alpha},- 1}]}\big\rangle_{M,\acute M}}
\nonumber
\end{eqnarray}
The saddle-point equations (\ref{eq:D},\ref{eq:P}) remain unaltered, with
the $\langle\ldots\rangle_M$ averages given by equation (\ref{def:M}), but now the associated measure takes the new form
\begin{eqnarray}\label{eq:Mvc}
M[\Hv,\hat{\hv},\sigmav\vert\theta]&=&\exp\Big\{\rmi\hat{\hv}
\cdot[\Hv- \thetav]-  \rmi\sum_{\alpha}
\hat{D}_{\alpha}(\sigma_{\alpha},H_{\alpha})\\
&&\hspace*{-5mm} +  c \sum_{{\sigmav^\prime}}\int\!\mathrm d
\hat{\hv}^\prime~ P({\sigmav^\prime} ,\hat{\hv}^\prime)\big[
e^{- \rmi J\sum_{\alpha}[\hat{h}_{\alpha}
\delta_{\sigma^\prime_{\alpha},- 1}+
\hat{h}^\prime_{\alpha} \delta_{\sigma_{\alpha},-
1}]}\!- 1\big]\Big\}\nonumber
\end{eqnarray}
The only changes to the earlier theory that are induced by the introduction of
(\ref{eq:VCfield})
are in the imaginary arguments of the exponential function in
(\ref{eq:Avc}) and (\ref{eq:Mvc}). We can therefore derive the RS
version of the theory for VC dynamics simply by replacing
$\sigma_{\alpha}\rightarrow\delta_{\sigma_{\alpha},-1}$ and
$P(J_\ell)\to\delta(J_\ell\!-\! J)$ in
equations (\ref{eq:W}-\ref{eq:Ars1}) of
section \ref{section:RS}. This results in
\begin{eqnarray}
\hspace*{-20mm}
W[\{P\}]&=&\sum_{k\geq 0}\frac{c^k}{k!}e^{-
c}\int\!\prod_{\ell=1}^k\Big\{ \{\mathrm d
P_\ell\} W[\{P_l\}]\Big\}
\label{eq:Wvc}\\
\hspace*{-20mm}
&&
\hspace*{-20mm}
\times\prod_{\sigma \hat h}\delta\Big[ P(\sigma,\hat h)-
\frac{\int\! \mathrm d H d(\sigma,H)\rme^{\rmi\hat{h}[ H\!-\!
\theta]}\prod_{\ell=1}^k\Big\{ \sum_{\sigma_{\ell}}\int\!\mathrm d
\hat{h}_{\ell} P_\ell(\sigma_{\ell},\hat{h}_{\ell})\rme^{- \rmi
J[\hat{h}\delta_{\sigma_{\ell},- 1}+\hat{h}_{\ell}
\delta_{\sigma , -
1}]}
\Big\}}{Z[\{P_1,\ldots,P_k\}]}\Big]\hspace*{-5mm}
\nonumber
\end{eqnarray}
and, with the Fourier transforms $\hat{P}(\sigma|x)=\int\! \mathrm d \hat h
~P(\sigma, \hat h)\rme^{- \rmi\hat h x}$,
\begin{eqnarray}
D(s,h) &=&~ d(s,h)\sum_{k\geq 0}\frac{c^k}{k!}e^{-
c}\int\!\prod_{\ell=1}^k\Big\{ \{ \mathrm d
P_\ell\} W[\{P_\ell\}]\Big\}
\label{eq:DvcRS1}\\
&&
\times \frac{\prod_{\ell=1}^k\Big\{
\sum_{\sigma_{\ell}}
\hat{P}_\ell(\sigma_{\ell}|J\delta_{s, - 1})\Big\}\delta
[h- J\sum_{\ell=1}^k \delta_{\sigma_{\ell},- 1}-
\theta]}{Z[\{P_1,\ldots,P_k\}]}\nonumber
\end{eqnarray}
\begin{eqnarray}
\hspace*{-15mm}
A[s,s^\prime;h,h^\prime;\tilde s]&=&\sum_{k\geq
0}\frac{c^k}{k!}e^{- c}\int\!\prod_{\ell=1}^k\Big\{
\{ \mathrm d P_\ell\} W[\{P_\ell\}]\sum_{\sigma_{\ell}}
\hat{P}_\ell(\sigma_{\ell}|J\delta_{s^\prime,- 1})\Big\}
\nonumber
\\
\hspace*{-15mm}
&&\times
\sum_{m\geq 0}\frac{c^m}{m!}e^{-
c}\int\!\prod_{r=1}^m\Big\{
\{ \mathrm d Q_r\} W[\{Q_r\}]\sum_{\sigma^{}_{r}}
\hat{P}_r(\sigma^{}_{r}|J\delta_{s,- 1})\Big\}
\label{eq:AvcRS1}\\
&&\nonumber \\
&&\times
\delta
[h^\prime\! -\! J\sum_{\ell=1}^k
\delta_{\sigma_{\ell},- 1}\!-\! \theta\!-\! J \delta_{s,- 1}]~
d(s^\prime,h^\prime)
\nonumber
\\
\hspace*{-15mm}
&&\times
\delta [ h\! -\! J\sum_{r=1}^m
\delta_{\sigma_{r},- 1}\!-\! \theta\!-\! J \delta_{s^\prime, - 1}\!-\! J\tilde s]~
 d(s,h-J\tilde{s})
 \nonumber\\
 \hspace*{-15mm}
%normalization of A
&&\times \left[
\sum_{\sigma\sigma^\prime}\prod_{\ell=1}^k\Big\{
\sum_{\sigma_{\ell}}
\hat{P}_\ell(\sigma_{\ell}|J\delta_{\sigma,- 1})\Big\} d\big(\sigma,J\sum_{\ell=1}^k
\delta_{\sigma_{\ell},- 1}\!+\! \theta\!+\! J \delta_{\sigma^\prime,- 1}\big)\nonumber
\right.
\\
\hspace*{-15mm}
&&\left.\times \prod_{r=1}^m\Big\{
\sum_{\sigma^{}_{r}}
\hat{P}_r(\sigma^{}_{r}|
J\delta_{\sigma^\prime,- 1})\Big\} d\big(\sigma^\prime,J\sum_{r=1}^m
\delta_{\sigma_{r},- 1}\!+\! \theta\!+\! J \delta_{\sigma,- 1}\big)\right]^{- 1}\nonumber
\end{eqnarray}
As before we may switch to a measure defined directly on the relevant Fourier transforms, which in the case of VC simplifies further
due to the uniform bonds $J$:
\begin{eqnarray}
\hspace*{-15mm}
\tilde W[\{\hat P\}]&=&\int\!\left\lbrace \mathrm d P\right\rbrace
W[\{P\}]\prod_{\sigma \sigma^\prime} \delta\Big[ \hat P(\sigma \vert
J\delta_{\sigma^\prime,- 1})- \int\! \mathrm d \hat h
~P(\sigma, \hat h)\rme^{- \rmi\hat h J\delta_{\sigma^\prime,-
1}}\Big]
\label{def:WvcFT}
\end{eqnarray}
Our RS  equations now acquire the following form:
\begin{eqnarray}
\hspace*{-20mm}
\tilde W[\{\hat P\}]&=& \sum_{k\geq 0}\frac{c^k}{k!}e^{-
c}\int\prod_{\ell=1}^k\Big\{
\{\mathrm d \hat P_\ell\} \tilde W[\{\hat P_\ell\}]\Big\}
\label{eq:WvcFT} \\
\hspace*{-20mm}
&&\hspace*{-17mm}
\times\prod_{\sigma \sigma^\prime}\delta\Big[ \hat P(\sigma \vert
J\delta_{\sigma^\prime\!,- 1})-\!
\frac{\prod_{\ell=1}^k\Big\{\! \sum_{\sigma_{\ell}}\! \hat
P_\ell(\sigma_{\ell}\vert J\delta_{\sigma, - 1})\Big\}
d(\sigma,J\sum_{\ell=1}^k\! \delta_{\sigma_\ell,- 1}\!+\! \theta\!+\!
J\delta_{\sigma^\prime,- 1})}{Z[\{\hat P_1,\ldots,\hat
P_k\}]}\Big]\nonumber
\\
\hspace*{-20mm}
D(s,h) &=&~d(s,h) \sum_{k\geq 0}\frac{c^k}{k!}e^{- c}\int\prod_{\ell=1}^k\Big\{
\{\mathrm d \hat P_\ell\} \tilde W[\{\hat P_\ell\}]\Big\}
\label{eq:DvcRS2}
\\
\hspace*{-20mm}
&&\hspace*{-0mm}
\times \frac{\prod_{\ell=1}^k\Big\{\!
\sum_{\sigma_{\ell}}\! \hat P_\ell(\sigma_{\ell}\vert J \delta_{s,-
1})\Big\} \delta
[h- J\sum_{\ell=1}^k \delta_{\sigma_{\ell},- 1}-
\theta]}{Z[\{\hat P_1,\ldots,\hat P_k\}]}\nonumber
\end{eqnarray}
and
\begin{eqnarray}
\hspace*{-15mm}
A[s,s^\prime;h,
h^\prime;\tilde s]&=&\sum_{k\geq 0}\frac{c^k}{k!}e^{- c}\int\!\prod_{\ell=1}^k\Big\{
\{\mathrm d \hat P_\ell\} \tilde W[\{\hat P_\ell\}]
\sum_{\sigma_{\ell}}\hat
P_\ell(\sigma_{\ell}\vert J\delta_{s^\prime,- 1})\Big\}
\label{eq:AvcRS2}
\\
\hspace*{-15mm}
&&\times\sum_{m\geq 0}\frac{c^m}{m!}e^{-
c}\int\!\prod_{r=1}^m\Big\{
\{\mathrm d \hat Q_r\} \tilde W[\{\hat Q_r\}]
\sum_{\sigma_{r}}\hat
Q_r(\sigma_{r}\vert J\delta_{s,- 1})\Big\}
\nonumber
 \\
 \hspace*{-15mm}
&&
\times
\delta
[h^\prime\! - J\sum_{\ell=1}^k
\delta_{\sigma_{\ell},- 1}- \theta- J \delta_{s,- 1}]
~d(s^\prime,h^\prime)
\nonumber
\\
\hspace*{-15mm}
&& \times
\delta [ h - J\sum_{r=1}^m
\delta_{\sigma_{r},- 1}- \theta- J \delta_{s^\prime,- 1}- J\tilde s]~ d(s,h-J\tilde{s})
\nonumber\\
%normalization of A
\hspace*{-15mm}
&&\times \left[\sum_{\sigma\sigma^\prime}
\prod_{\ell=1}^k\Big\{ \sum_{\sigma_{\ell}}\hat
P_\ell(\sigma_{\ell}\vert J\delta_{\sigma,- 1})\Big\}  d\big(\sigma,J\sum_{\ell=1}^k
\delta_{\sigma_{\ell},- 1}\!+\! \theta\!+\! J \delta_{\sigma^\prime\!,- 1}\big)\right.
\nonumber\\
\hspace*{-15mm}
&&
\left.\times \prod_{r=1}^m\Big\{
\sum_{\sigma_{r}}
\hat Q_r(\sigma_{r}\vert J\delta_{\sigma^\prime,- 1})\Big\}  d\big(\sigma^\prime\!,J\sum_{r=1}^m
\delta_{\sigma_{r},- 1}\!+\! \theta\!+\! J \delta_{\sigma,- 1}\big)  \right]^{- 1}\nonumber
\end{eqnarray}
 Compared to the more general expression (\ref{eq:Wft}), in the VC case (\ref{eq:WvcFT}) the dimensionality of our problem has
  been reduced drastically, as $\tilde W$ is now a
functional on the space of $2\times2$ matrices $\hat{P}(\sigma|J\delta_{\sigma^\prime,-1})$. Furthermore, the solutions of (\ref{eq:DvcRS2},\ref{eq:AvcRS2}) are of the following form, which is expected on physical
grounds (given the non-random bonds $J$ in VC):
\begin{eqnarray}
D(s,h)&=&\sum_{n\geq0}P(s,n)~\delta (h- Jn-
\theta)\label{def:Psn}
\\
A[s,s^\prime;h,h^\prime;\tilde s]&=&\sum_{n,\acute n\geq 0}A[s,s^\prime
;n,n^\prime]~\delta [h^\prime\! -\! Jn^\prime\!-\! \theta\!-\! J
\delta_{s,- 1}]
\nonumber
\\
&&\hspace*{20mm}\times
\delta [ h\! -\! Jn\!-\! \theta\!-\! J
\delta_{s^\prime\!,- 1}- J\tilde s]
\label{def:Asn}
\end{eqnarray}
where $P(s,n)$ and $A[s,
s^\prime;n,n^\prime]$ (with $s,s^\prime\in\{-1,1\}$ and $n,n^\prime\in \{0,1,2,\ldots\}$) are solved from
\begin{eqnarray}
P(s,n)&=&\sum_{k\geq 0}\frac{c^k}{k!}e^{- c}\int\!\prod_{\ell=1}^k\Big\{
\{\mathrm d \hat P_\ell\} \tilde W[\{\hat P_\ell\}]\Big\}
\label{eq:Psn}\\
&&\times \frac{\prod_{\ell=1}^k\Big\{ \sum_{\sigma_{\ell}} \hat
P_\ell(\sigma_{\ell}\vert J \delta_{s,- 1})\Big\}
d(s,Jn\!+\! \theta)\delta_{n,\sum_{\ell=1}^k
\delta_{\sigma_{\ell},-
1}}}{\sum_{\sigma}\prod_{\ell=1}^k\Big\{ \sum_{\sigma_{\ell}} \hat
P_\ell(\sigma_{\ell}\vert J \delta_{\sigma,- 1})\Big\}
d(\sigma,J\sum_{\ell=1}^k \delta_{\sigma_{\ell},- 1}+
\theta\big)}\nonumber
\end{eqnarray}
and
\begin{eqnarray}
\hspace*{-15mm}
A[s,s^\prime;n,n^\prime]&=&\sum_{k\geq 0}\frac{c^k}{k!}e^{- c}\int\!\prod_{\ell=1}^k\Big\{
\{\mathrm d \hat P_\ell\} \tilde W[\{\hat P_\ell\}]\sum_{\sigma_{\ell}}\hat
P_\ell(\sigma_{\ell}\vert J\delta_{s^\prime,- 1})\Big\}
\label{eq:Asn}\\
\hspace*{-15mm}
&&\times
\sum_{m\geq 0}\frac{c^m}{m!}e^{-
c}\int\prod_{r=1}^m\Big\{
\{\mathrm d \hat Q_r\} \tilde W[\{\hat Q_r\}]\sum_{\sigma_{r}}\hat
Q_r(\sigma_{r}\vert J\delta_{s,- 1})\Big\}\nonumber \\
\hspace*{-15mm}
&&
\times d(s^\prime,Jn^\prime\!+ \!\theta\!+ \!J \delta_{s,- 1})~\delta_{n^\prime,\sum_{\ell=1}^k
\delta_{\sigma_{\ell},- 1}}\nonumber\\
\hspace*{-15mm}
&&
\times d(s,Jn\!+\! \theta\!+\! J \delta_{s^\prime,- 1})~\delta_{n,\sum_{r=1}^m
\delta_{\sigma_{r},- 1}}\nonumber
\\
\hspace*{-15mm}
%normalization of A
&&\times \left[
\sum_{\sigma \sigma^\prime}
\prod_{\ell=1}^k\Big[ \sum_{\sigma_{\ell}}\hat
P_\ell(\sigma_{\ell}\vert J\delta_{\sigma,- 1})\Big]  d\big(\sigma,J\sum_{\ell=1}^k
\delta_{\sigma_{\ell},- 1}\!+\! \theta\!+\! J \delta_{\sigma^\prime\!,- 1}\big)\nonumber
\right.
\\
\hspace*{-15mm}
&&\left.
\times\prod_{r=1}^m\Big[ \sum_{\sigma_{r}}
\hat Q_r(\sigma_{r}\vert J\delta_{\sigma^\prime,- 1})\Big]  d\big(\sigma^\prime\!,J\sum_{r=1}^m
\delta_{\sigma_{r},- 1}\!+\! \theta\!+\! J \delta_{\sigma,- 1}\big)  \right]^{- 1}\nonumber
\end{eqnarray}
The simplified form of the kernels (\ref{def:Psn},\ref{def:Asn}) subsequently
allows us to transform the main dynamical equation (\ref{eq:diffusion}), which is a PDE, into the following
system of ordinary differential equations (see \ref{section:ODE} for details):
\begin{eqnarray}
 \frac{\mathrm d}{\mathrm d t}P(s,0)&=& \frac{1}{2}\left [1\!+\! s\tanh[\beta\theta]\right
]P(\!-\! s,0)- \frac{1}{2}\left [1\!-\!
s\tanh[\beta\theta]\right
]P(s,0)\nonumber\\
&&+ \frac{1}{2}c\sum_{n^\prime\geq 0}
[1\!+\! \tanh[\beta Jn^\prime \!+\! \beta\theta\!+\! \beta J \delta_{s,- 1}]]A[s,\!-\! 1;0,n^\prime]\nonumber\\
&&-\frac{1}{2}c\sum_{n^\prime\geq 0} [1\!-\! \tanh[\beta
J\acute n\!+\! \beta\theta\!+\! \beta J \delta_{s,-
1}]]A[s,1;0,n^\prime]
\label{eq:ODE0}
\end{eqnarray}
whereas for $n>0$ we have
\begin{eqnarray}
\hspace*{-15mm}
 \frac{\mathrm d}{\mathrm d t}P(s,n)&=& \frac{1}{2}\left [1\!+\! s\tanh[\beta Jn\!+\! \beta\theta]\right
]P(\!-\! s,n)- \frac{1}{2}\left [1\!-\! s\tanh[\beta
Jn\!+\! \beta\theta]\right
]P(s,n)\nonumber
\\
&&+ \frac{1}{2}c\sum_{n^\prime\geq 0}
[1\!-\! \tanh[\beta Jn^\prime \!+\! \beta\theta\!+\! \beta J \delta_{s,- 1}]]A[s,1;n\!-\!1,n^\prime]\nonumber\\
& &+ \frac{1}{2}c\sum_{n^\prime\geq 0}
[1\!+\! \tanh[\beta J n^\prime \!+\! \beta\theta\!+\! \beta J \delta_{s,- 1}]]A[s,\!-\!1;n,n^\prime]\nonumber\\
&& -\frac{1}{2}c\sum_{n^\prime\geq 0}
[1\!-\! \tanh[\beta Jn^\prime\!+\! \beta\theta\!+\! \beta J \delta_{s,- 1}]]A[s,1;n,n^\prime]\nonumber\\
&& -\frac{1}{2}c\sum_{n^\prime\geq 0} [1\!+\! \tanh[\beta
Jn^\prime \!+\! \beta\theta \!+\! \beta J \delta_{s,-
1}]]A[s,\!-\! 1;n\!-\! 1,n^\prime]
\label{eq:ODE}
\end{eqnarray}
Equations (\ref{eq:WvcFT}) and (\ref{eq:Psn}-\ref{eq:ODE}) are the final results of our dynamical replica analysis.
They can be solved numerically, using population dynamics for the functional saddle-point equations (see \ref{section:population} for
 details) and
any standard method for the system of ordinary differential equations.
If we allow for temperature adaptation, viz. $\beta\to\beta(t)$, and restrict ourselves to those cooling protocols where $\beta(t)$ changes only on $\order(N^0)$ time scales, we may simply make the replacement $\beta\to\beta(t)$ in the above equations.

\section{Tests of the VC theory against numerical simulations}
\label{section:simulations}
 To test our theoretic predictions for the evolution of observables in the Glauber algorithm with stochastic cooling running on the VC problem,
 we compare the results of solving numerically the system of dynamical equations (\ref{eq:ODE}) with the results of numerical simulations.
 We solve (\ref{eq:ODE}) using a simple first-order Euler method, i.e. we iterate
 the iteration
\begin{equation}
P_{\ell+1}(s,n)=P_{\ell}(s,n)+ h\Gamma\left[s,n;P_{\ell}(\ldots);A_{\ell}[\ldots]\right],~~~~~~t_{\ell}=\ell h\label{eq:ODEdiscr}
\end{equation}
where $n\in\{0,1,\ldots, L(c)\}$ and $\Gamma[\ldots]$ is a short-hand for the right-hand side of (\ref{eq:ODE0},\ref{eq:ODE}).
Here $L(c)$ denotes a suitable cut-off value that increases monotonically with the average connectivity $c$ in (\ref{eq:P(c)}), and $0<h\ll 1$. At each discrete time-step $\ell$ of this iteration we solve the RS equations (\ref{eq:WvcFT},\ref{eq:Psn}) via a population dynamics algorithm (see section \ref{section:population}) and compute the kernel (\ref{eq:Asn}).
 Solving (\ref{eq:ODE}) requires a significant numerical effort,  the bulk of which is devoted to solving equations (\ref{eq:WvcFT},\ref{eq:Psn}), where the  computation of a typical object (\ref{eq:Pupdate}) requires typically $O(2^c)$ basic operations.
Although we expect that for sufficiently small $h$ the changes in the statistical properties of the population between consecutive iterative time-steps in (\ref{eq:ODEdiscr}) are small, the fact that the running time of the algorithm (\ref{eq:ODEdiscr}) grows exponentially with $c$ restricts the scope of simulation experiments.
For each choice of control parameters our experimental protocol has been the following.
First we generate a large random Poissonnian graph with the required connectivity $c$ . Then we run the algorithm (\ref{eq:algorithm}) with the local fields (\ref{eq:VCfield}), from an initial spin configuration where the individual spins are drawn randomly and independently from the distribution (\ref{eq:Prob0}). We then let the system evolve according to the Glauber
algorithm, but with the temperature $T(t)$ decreasing in stages (to achieve stochastic cooling), while we record the evolution of two macroscopic order parameters, being the fraction $x$ of covered vertices
\begin{equation}
x(\sigmav)=\frac{1}{N}\sum_i\delta_{\sigma_i,1}\label{eq:density}
\end{equation}
and the energy density $E$, which is proportional to the fraction of uncovered edges,
\begin{equation}
E(\sigmav)=\frac{1}{N}\sum_{i<j}c_{ij}\delta_{\sigma_i,-1}\delta_{\sigma_j,-1}\label{eq:uncovedges}
\end{equation}
For a state $\sigmav$ to represent an acceptable {\em vertex cover} it must have $E(\sigmav)=0$. For such a cover to be
{\em minimal} we want in addition $x(\sigmav)$ to be as small as possible.

In simulated annealing one starts a Monte Carlo dynamics at a high temperature $T(0)$, and then lowers it slowly in stages, allowing the system to
equilibrate effectively along the way. The objective is for the algorithm not to get stuck in states that are only locally but not globally optimal. Determining the best cooling schedule $T(t)$ for achieving this, however, is highly nontrivial; furthermore,  equilibration times
 in VC-type optimization problems can scale exponentially in the system size. Here we did not attempt to optimize the cooling protocol but
 focused on the VC dynamics for simple step-wise temperature reductions.
  Our numerical simulations where carried out on random graphs with $N=10,000$ vertices, with average connectivities $c\in\{0.5,1,1.5,2,2.5,3,3.5\}$. The values of the parameters $J=1$, $\theta=-0.99$, the initial covered fraction  $x_0=N^{-1}\sum_i\delta_{\sigma_i(0),1}=0.9$ and the temperatures $T\in\{2,1,0.5,0\}$ (reduced in steps) were identical in all simulations. The initial conditions for  (\ref{eq:ODE}) and the population dynamics were computed via equations (\ref{eq:P0(s,n)}-\ref{eq:d0}) of \ref{section:Initial conditions}. The size of the population was $\mathcal{N}=10,000$ and the number of iterations typically needed for the population dynamics to converge was of order $10\mathcal{N}$.

  \begin{figure}[t]
\vspace*{5mm} \hspace*{-9mm} \setlength{\unitlength}{0.40mm}
\begin{picture}(350,100)

   \put(0,0){\includegraphics[height=110\unitlength,width=150\unitlength]{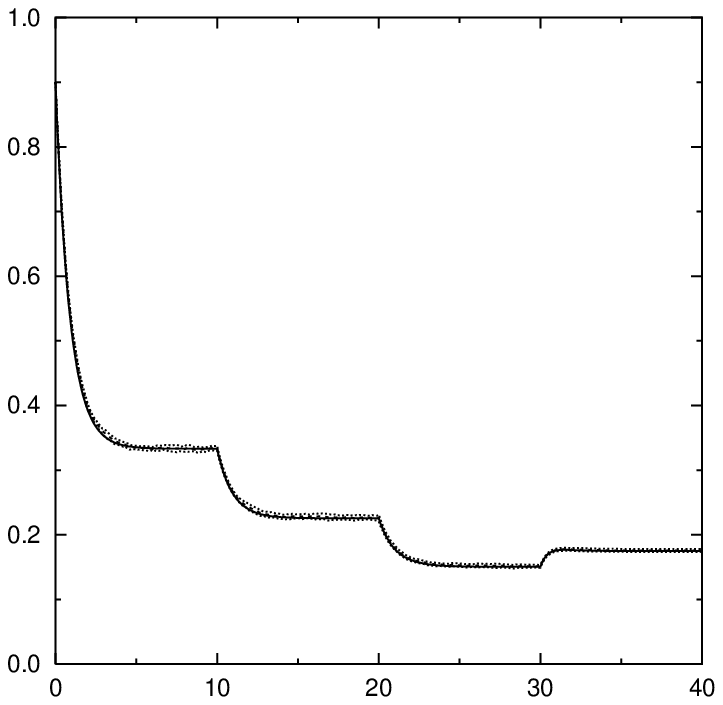}}
     \put(-3,65){\small\here{$x$}}
 \put(115,0){\includegraphics[height=110\unitlength,width=150\unitlength]{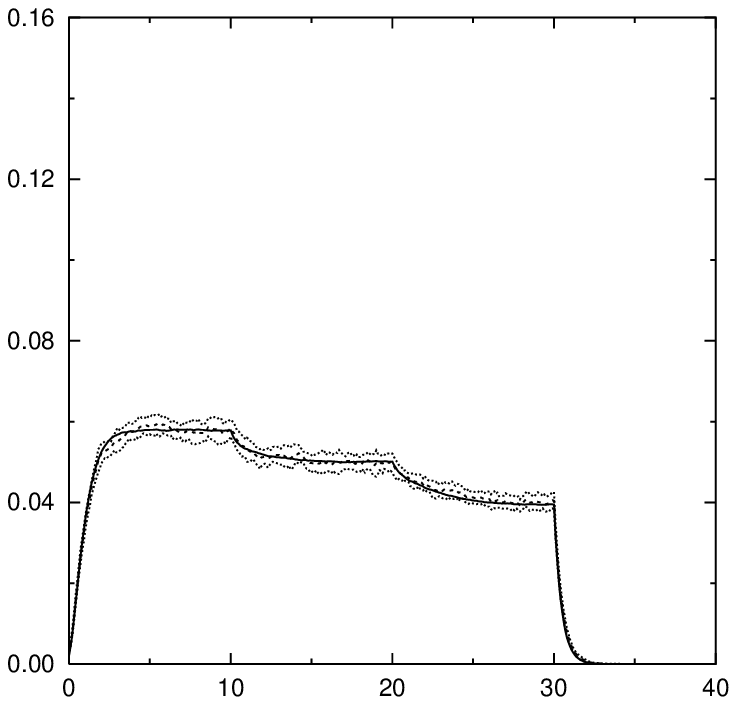}}
 \put(117,65){\small\here{$E$}}
 \put(235,-2.5){\includegraphics[height=112\unitlength,width=150\unitlength]{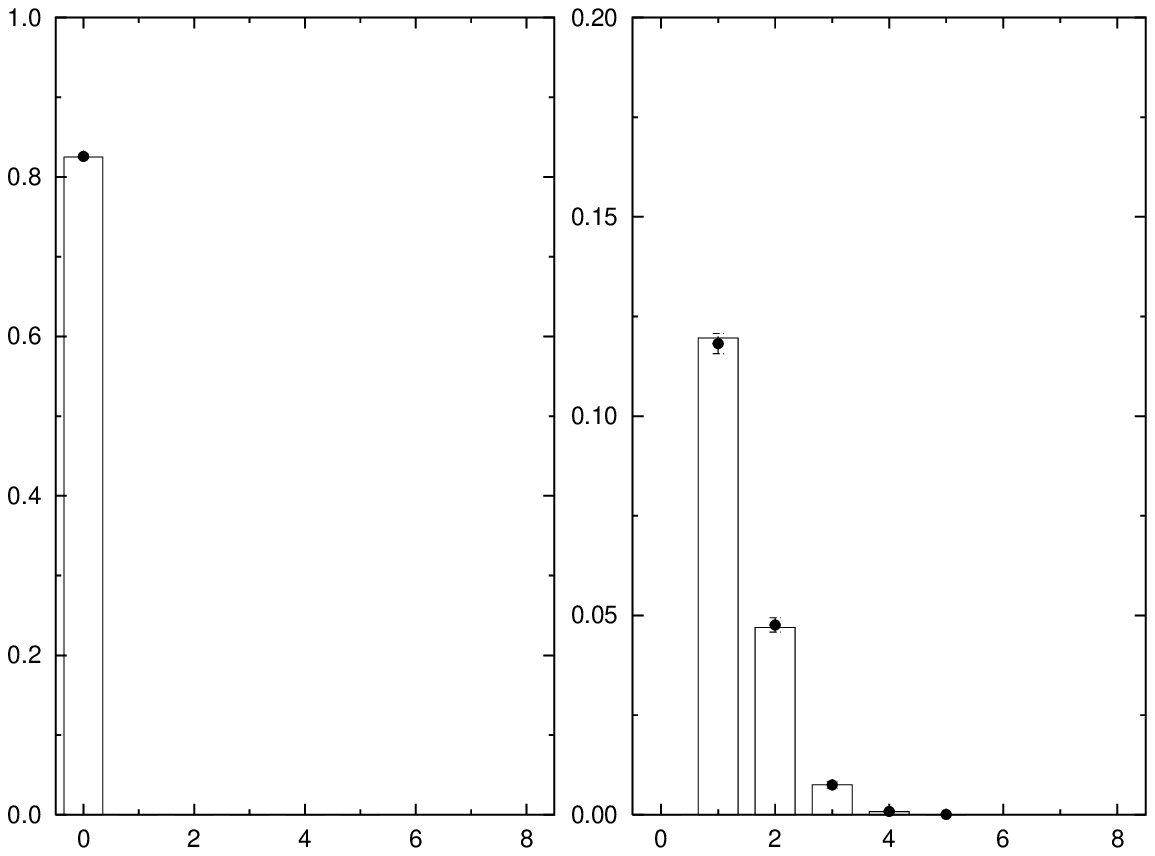}}
  \put(58,-12){\small $t$}    \put(175,-12){\small $t$}     \put(270,-12){\small $n$} \put(330,-12){\small $n$}
\put(277,95){\here{\small $P(-1,n)$}} \put(342,95){\here{\small $P(1,n)$}}

\end{picture}
 \vspace*{2mm}
\caption{Left and middle: evolution of the fraction $x$ and the energy density $E$ in the VC algorithm with simulated annealing,
 for $c=0.5$, $J=1.0$ and $\theta=-0.99$.
Time is measured in iterations per spin. Solid lines: RS theory. Dashed and dotted lines: average and average plus/minus standard deviation
as measured over 100 simulation runs in systems with $N=10^4$ spins. The annealing schedule had four stages: (i) $T=2$ for $t\in[0,10]$,
 (ii) $T=1$ for $t\in[10,20]$, (iii) $T=0.5$ for $t\in[20,30]$, (iv) $T=0$ for $t\in[30,40]$. Right: histograms (RS theory) of the two field
 distributions $P(\pm 1,n)$ at $t=40$, together with the corresponding simulation measurements (markers with error bars).} \label{fig:c05}
\end{figure}
%---------------------------------------figures----------------------------------------------
 \begin{figure}[t]
\vspace*{5mm} \hspace*{-9mm} \setlength{\unitlength}{0.40mm}
\begin{picture}(350,100)

 \put(0,0){\includegraphics[height=110\unitlength,width=150\unitlength]{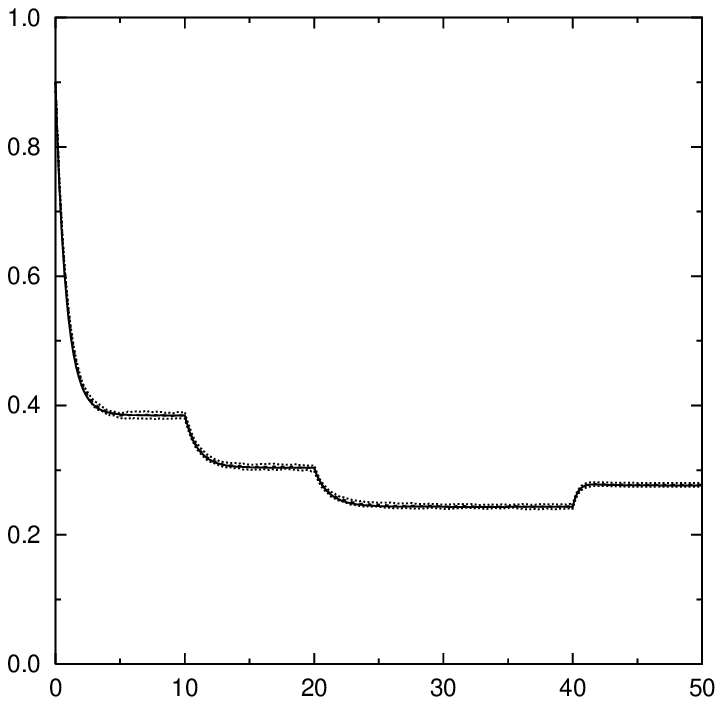}}
 \put(-3,65){\small\here{$x$}}
 \put(115,0){\includegraphics[height=110\unitlength,width=150\unitlength]{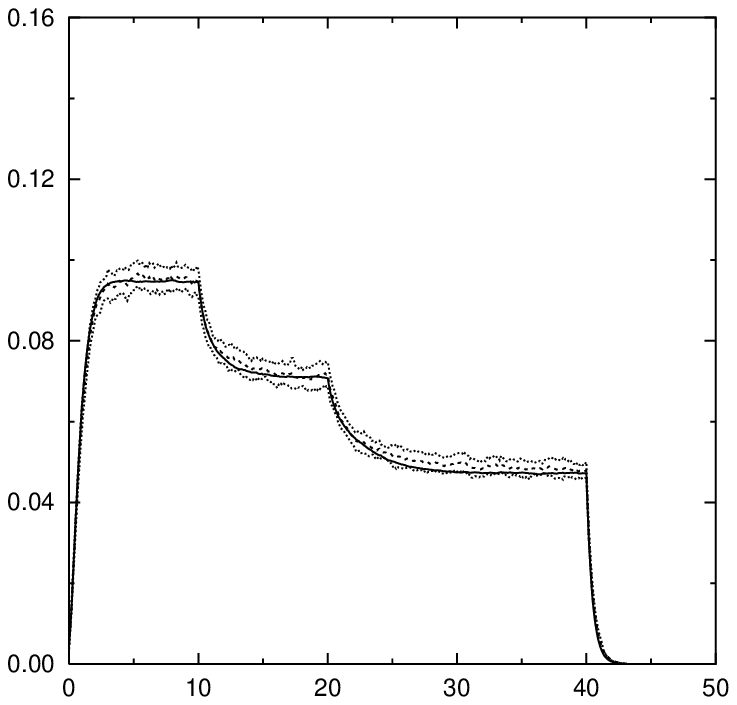}}
 \put(117,65){\small\here{$E$}}
  \put(235,-2.5){\includegraphics[height=112\unitlength,width=150\unitlength]{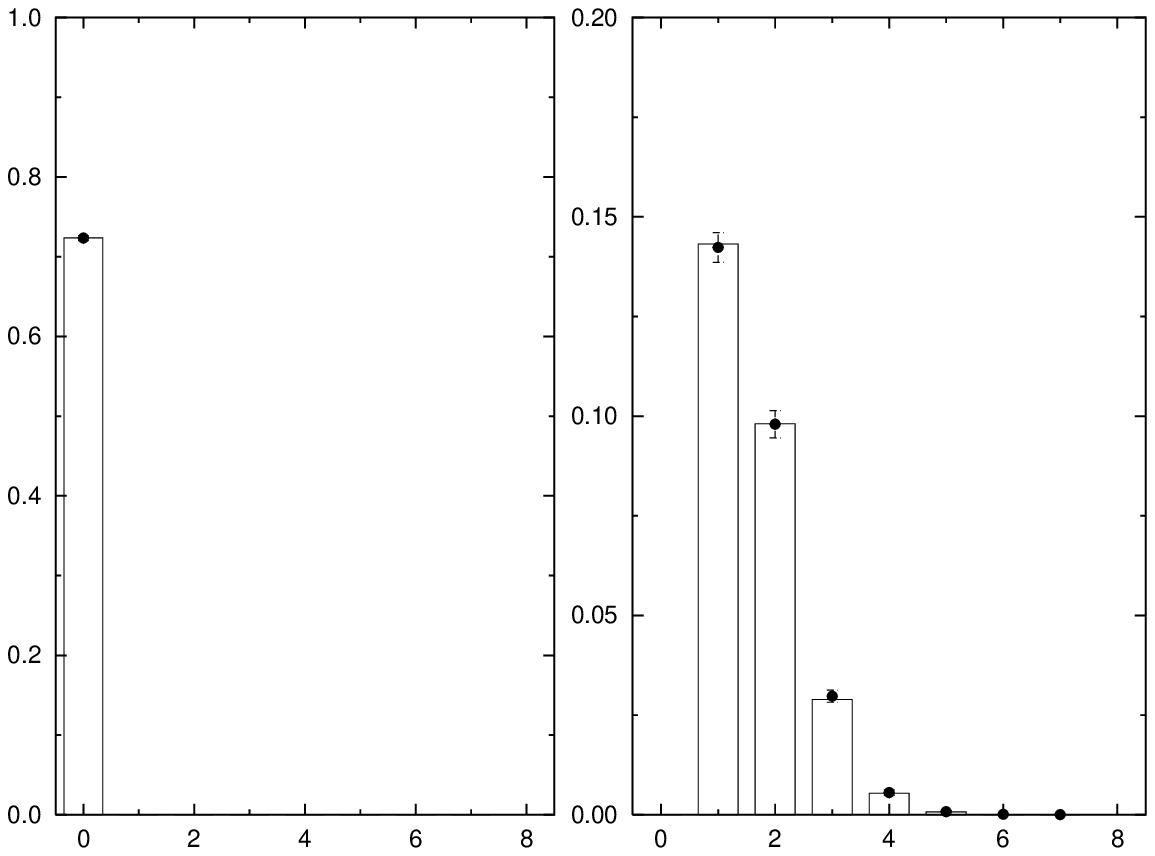}}
  \put(58,-12){\small $t$}    \put(175,-12){\small $t$}     \put(270,-12){\small $n$} \put(330,-12){\small $n$}
\put(277,95){\here{\small $P(-1,n)$}} \put(342,95){\here{\small $P(1,n)$}}

\end{picture}
 \vspace*{2mm}
\caption{
Left and middle: evolution of the fraction $x$ and the energy density $E$ in the VC algorithm with simulated annealing,
 for $c=1$, $J=1.0$ and $\theta=-0.99$.
Time is measured in iterations per spin. Solid lines: RS theory. Dashed and dotted lines: average and average plus/minus standard deviation
as measured over 100 simulation runs in systems with $N=10^4$ spins. The annealing schedule had four stages: (i) $T=2$ for $t\in[0,10]$,
 (ii) $T=1$ for $t\in[10,20]$, (iii) $T=0.5$ for $t\in[20,40]$, (iv) $T=0$ for $t\in[40,50]$. Right: histograms (RS theory) of the two field
 distributions $P(\pm 1,n)$ at $t=50$, together with the corresponding simulation measurements (markers with error bars).
} \label{fig:c1}
\end{figure}
 \begin{figure}[t]
\vspace*{5mm} \hspace*{-9mm} \setlength{\unitlength}{0.40mm}
\begin{picture}(350,100)

 \put(0,0){\includegraphics[height=110\unitlength,width=150\unitlength]{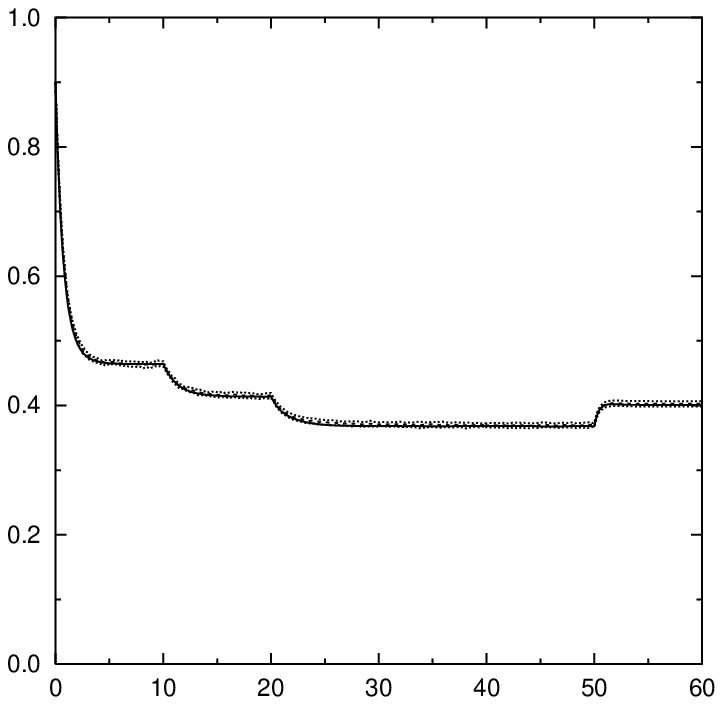}}
 \put(-3,65){\small\here{$x$}}
 \put(115,0){\includegraphics[height=110\unitlength,width=150\unitlength]{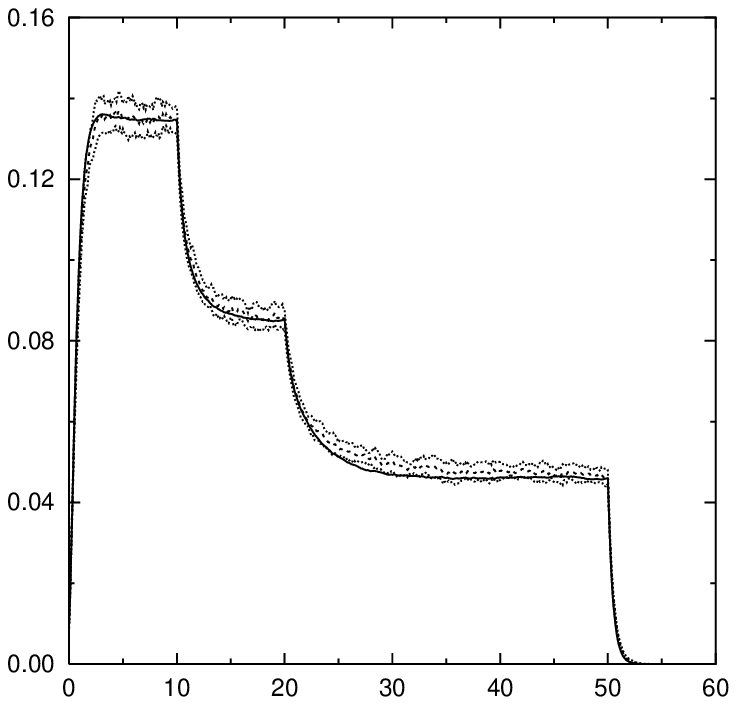}}
 \put(117,65){\small\here{$E$}}
 \put(235,-2.5){\includegraphics[height=112\unitlength,width=150\unitlength]{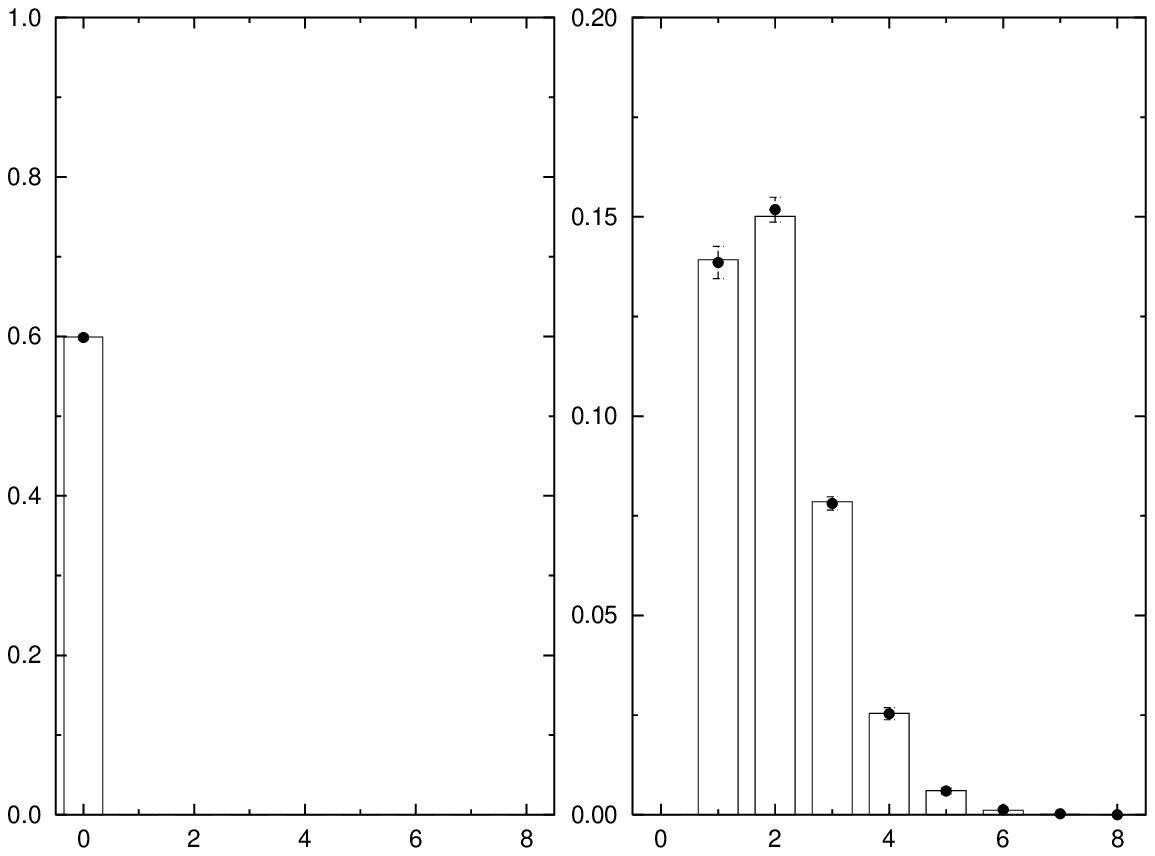}}
  \put(58,-12){\small $t$}    \put(175,-12){\small $t$}     \put(270,-12){\small $n$} \put(330,-12){\small $n$}
\put(277,95){\here{\small $P(-1,n)$}} \put(342,95){\here{\small $P(1,n)$}}

\put(60,62){\epsfysize=40\unitlength\epsfbox{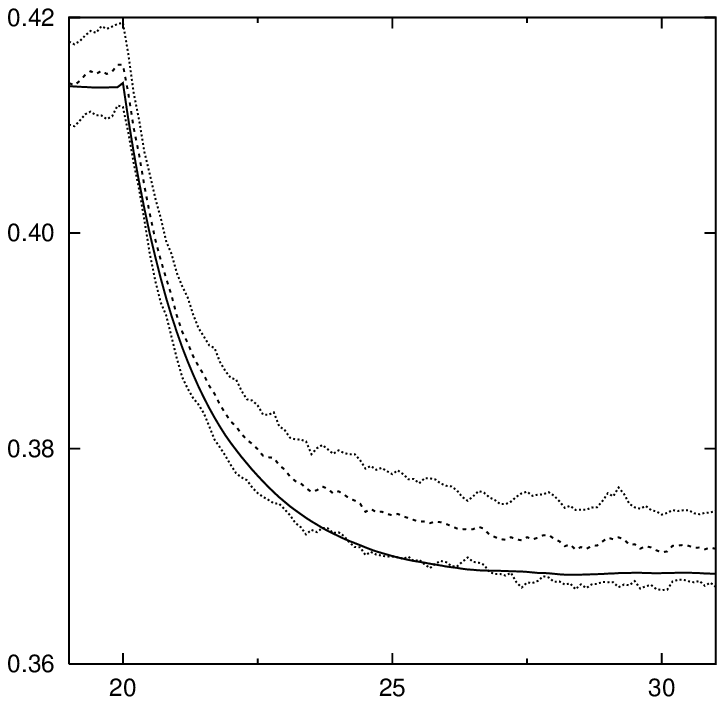}}

\end{picture}
 \vspace*{2mm}
\caption{
Left and middle: evolution of the fraction $x$ and the energy density $E$ in the VC algorithm with simulated annealing,
 for $c=2$, $J=1.0$ and $\theta=-0.99$.
Time is measured in iterations per spin. Solid lines: RS theory. Dashed and dotted lines: average and average plus/minus standard deviation
as measured over 100 simulation runs in systems with $N=10^4$ spins. The annealing schedule had four stages: (i) $T=2$ for $t\in[0,10]$,
 (ii) $T=1$ for $t\in[10,20]$, (iii) $T=0.5$ for $t\in[20,50]$, (iv) $T=0$ for $t\in[50,60]$.
 The inset in the left figure shows an enlargement of the region $t\in[20,30]$, where
 the largest deviation between theory and simulation for $x$ is observed.
 Right: histograms (RS theory) of the two field
 distributions $P(\pm 1,n)$ at $t=60$, together with the corresponding simulation measurements (markers with error bars).
} \label{fig:c2}
\end{figure}
 \begin{figure}[t]
\vspace*{5mm} \hspace*{-9mm} \setlength{\unitlength}{0.40mm}
\begin{picture}(350,100)

   \put(0,0){\includegraphics[height=110\unitlength,width=150\unitlength]{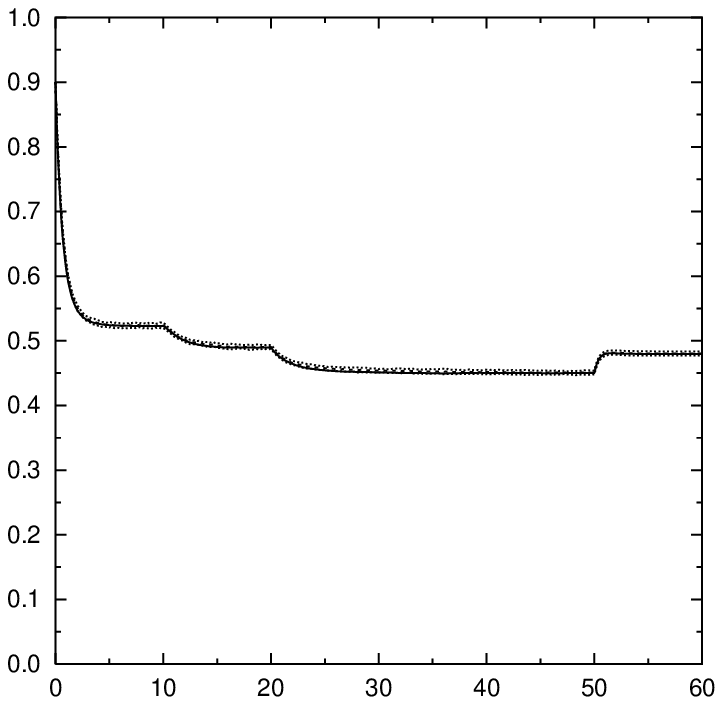}}
     \put(-3,65){\small\here{$x$}}
 \put(115,0){\includegraphics[height=110\unitlength,width=150\unitlength]{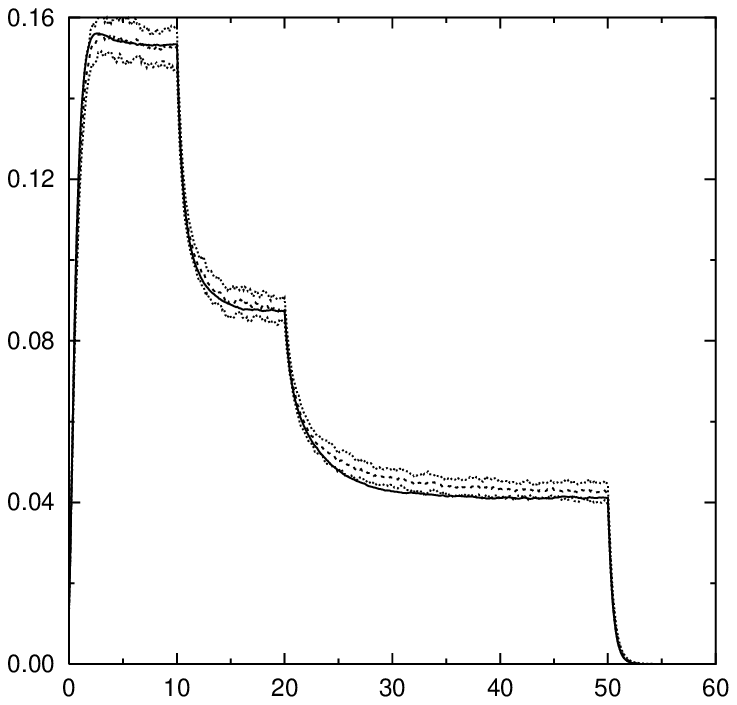}}
 \put(117,65){\small\here{$E$}}
  \put(235,-2.5){\includegraphics[height=112\unitlength,width=150\unitlength]{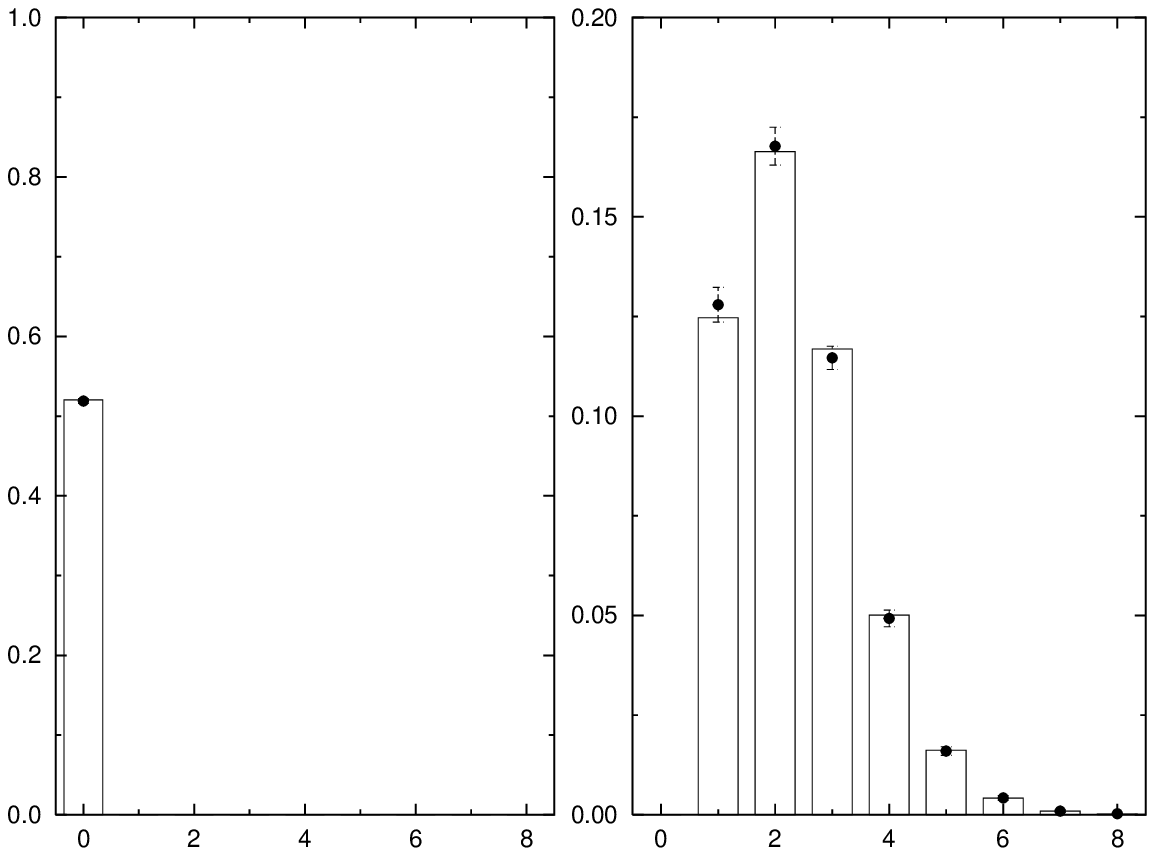}}
  \put(58,-12){\small $t$}    \put(175,-12){\small $t$}     \put(270,-12){\small $n$} \put(330,-12){\small $n$}

\put(277,95){\here{\small $P(-1,n)$}}
\put(342,95){\here{\small $P(1,n)$}}

\put(60,62){\epsfysize=40\unitlength\epsfbox{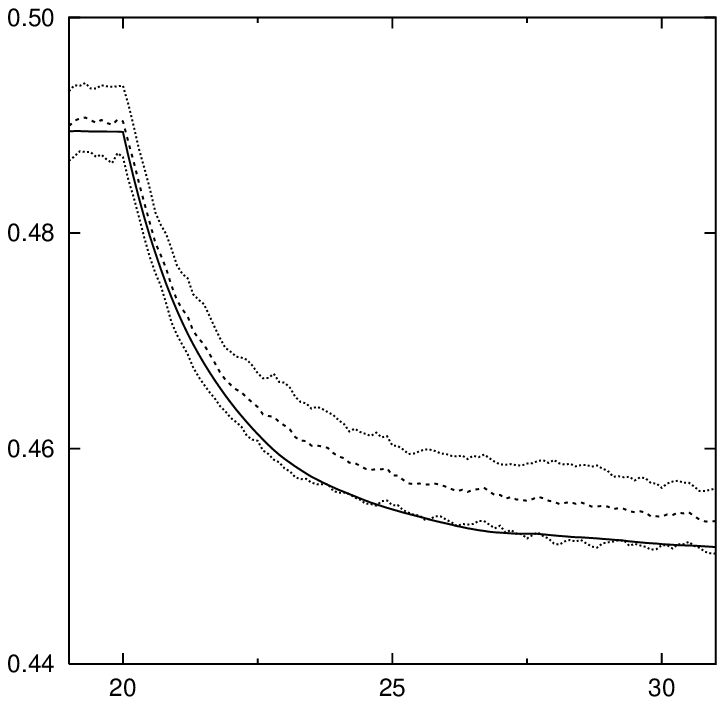}}

\end{picture}
 \vspace*{2mm}
 \caption{
 Left and middle: evolution of the fraction $x$ and the energy density $E$ in the VC algorithm with simulated annealing,
 for $c=3$, $J=1.0$ and $\theta=-0.99$.
Time is measured in iterations per spin. Solid lines: RS theory. Dashed and dotted lines: average and average plus/minus standard deviation
as measured over 100 simulation runs in systems with $N=10^4$ spins. The annealing schedule had four stages: (i) $T=2$ for $t\in[0,10]$,
 (ii) $T=1$ for $t\in[10,20]$, (iii) $T=0.5$ for $t\in[20,50]$, (iv) $T=0$ for $t\in[50,60]$.
 The inset in the left figure shows an enlargement of the region $t\in[20,30]$, where
 the largest deviation between theory and simulation for $x$ is observed.
 Right: histograms (RS theory) of the two field
 distributions $P(\pm 1,n)$ at $t=60$, together with the corresponding simulation measurements (markers with error bars).
 } \label{fig:c3}
\end{figure}

\begin{figure}[t]
\vspace*{1mm} \hspace*{35mm} \setlength{\unitlength}{0.50mm}
\begin{picture}(100,100)
\put(0,0){\epsfysize=100\unitlength\epsfbox{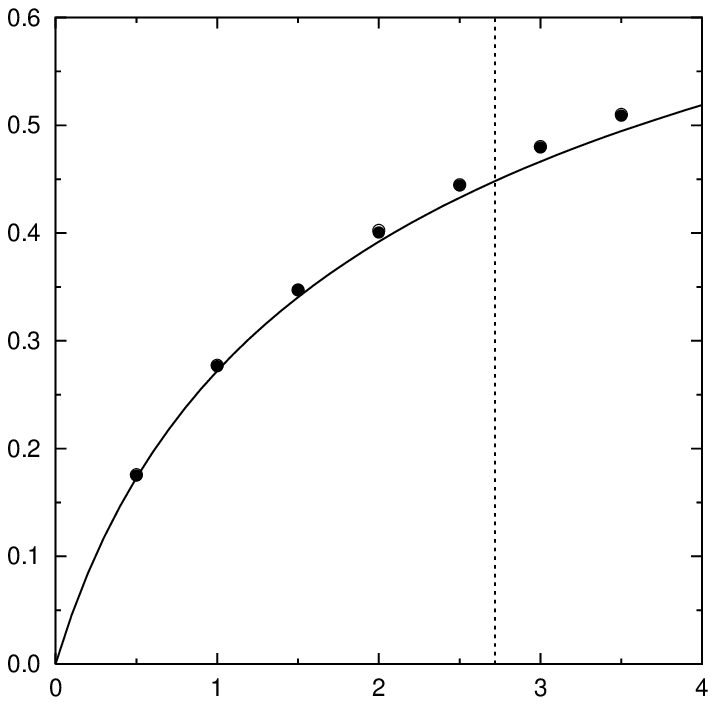}}
\put(55,-10){\here{\small $c$}}
\put(-9,55){\here{\small $x(c)$}}

\put(40,15) {\rm RS} \put(82,15){\rm RSB}

\end{picture}
\vspace*{8mm} \caption{The fraction $x(c)$ of covered vertices in a minimal vertex cover as a function of the average connectivity $c$.
Solid line: prediction  of a static replica-symmetric calculation (exact for $c<e$; the value $c=e$ is shown as a vertical dashed line),
as obtained in \cite{VCising1}. Symbols:
the predicted final fraction $x$ of covered vertices in the vertex cover obtained from the present dynamics (when we have arrived at $T=0$),
according to the RS dynamical replica method (which agrees perfectly with the simulations).} \label{fig:diagram}
\end{figure}

 In figures \ref{fig:c05}-\ref{fig:c3} we compare the data obtained in our numerical simulations for $c\in\{0.5,1,2,3\}$ with the results of solving  (\ref{eq:ODE}) numerically. We observe that the overall agreement between theory and simulations is excellent.
The RS theory also predicts correctly the joint spin-field statistics $P_t(s,n)$, see the right panels  in figures \ref{fig:c05}-\ref{fig:c3}; note the different vertical scales.
The deviations between theory and simulations are (as usual in DRT) confined to intermediate times, and limited to low temperatures in combination
 with high average connectivity, as shown in the insets of figures \ref{fig:c2} and \ref{fig:c3}; but even there they remain within the error bars of the simulation data.
Finally, in figure \ref{fig:diagram}, we compare our results for the fraction of covered vertices $x$ as measured at termination of the algorithm with the result (\ref{eq:treshold})
of equilibrium statistical mechanics, as obtained (within the replica-symmetry ansatz) in \cite{VCising1}.
Our data (markers, with virtually no difference between the observed values in simulations and the prediction of our dynamical theory) are seen to be
close to the equilibrium prediction (\ref{eq:treshold}) (solid curve), but they overestimate slightly the size of minimal vertex covers. This type of behaviour is not unusual in simulations, and suggests that more sophisticated annealing schemes must be used to achieve equilibration.
Slightly more unexpected is the fact that our RS theory exhibits a similar overestimation of $x$. This, in combination with the fact that the static
RS equations of \cite{VCising1} can be shown to constitute a stationary solution of our present dynamical replica equations,
suggests (at least to the left of the vertical dashed line, where replica symmetry
should hold)  that the time required for true equilibration diverges with $N$.
Both the static RS replica equations and the long time limit of the dynamical RS replica equations represent distinct
stationary solutions of the dynamical formalism, and the observed differences in $x(c)$ are manifestations of the non-commuting
of the limits $N\to\infty$ and $t\to\infty$.

\section{Discussion}\label{section:Discussion}
In this paper we have studied the sequential dynamics of finitely connected Ising spin models with random bonds, on Poissonnian random graphs.
Starting from the microscopic master equation we derived a  dynamic equation for
  the joint spin-field probability distribution, which is exact in the infinite system size limit, but not closed.
  We then followed the usual prescriptions and assumptions of dynamic replica theory \cite{DRTinf2} in order to close this equation.
  The result is a set of nontrivial coupled diffusion equations, in which the evaluation of the driving forces  requires the
  solution of a saddle-point problem at each instance of time. The latter saddle-point equations are of a functional nature, and are
   derived within the replica-symmetric (RS) ansatz; they can be solved numerically by a conventional population dynamics algorithm \cite{BetheSG}.

As a first application,  we have applied our dynamical theory to the dynamics of a simulated annealing algorithm (Glauber-type dynamics with a stepwise stochastic cooling schedule) when running to find a solution of the so-called minimal vertex cover (VC) problem \cite{VCising1},
 on finitely connected Poissonnian random graphs. In this problem the local fields are essentially integer-valued, which simplifies our dynamical equations. We have derived dynamic equation for the joint probability of spins and nonnegative integer fields. Upon solving the equations of our theory numerically and comparing the results with the outcome of numerical simulations of the algorithm, we find excellent agreement between theory and experiment.

When compared to e.g. the generating functional analysis method (GFA), the advantage of dynamical replica theory (DRT) is that, unlike GFA, it does not give an effective number of scalar order parameters that grows exponentially with time.
Although also in its present form\footnote{For parallel stochastic dynamics it can be shown that an exact formulation of DRT is possible
for any discrete spin model,
 with an effective number of scalar order parameters that grows at most linearly with time \cite{ExactDRT}.}, with the joint spin-field distribution as the core dynamical order parameter,
the DRT method is not exact, it is certainly much more accurate than e.g. any simple two-parameter theory \cite{DRTfc}, and it can be systematically improved further by increasing the order parameter set \cite{ApprSch}, although at a numerical cost.
We believe its wide applicability to be the main advantage of the dynamical theory presented in this paper.
The formalism can be extended relatively easily to include, for instance, directed or non-Poissonnian random graphs.

\section*{Acknowledgements}

It is our great pleasure to  thank  I P\'{e}rez-Castillo, JPL Hatchett, A Annibale and M Weigt for interesting and helpful discussions.

\appendix
\section{Averaging over disorder}\label{section:average}
In this section we give the details of the disorder averaging in equation (\ref{eq:A2}) that brings us to equation (\ref{eq:A3}).
Firstly, we rewrite slightly the term within the angular brackets, exploiting the symmetry of $c_{ij}J_{ij}$ under index permutations
$i\leftrightarrow j$:
\begin{eqnarray}
\hspace*{-20mm}
\bra \ldots\ket_{\{c_{ij}J_{ij}\}}&=&
\langle c_{12}\delta [ h - H_2^1+ 2J_{12}\tilde
s]\rme^{- \rmi\sum_{\alpha  i}\hat{h}_i^{\alpha}h_i
(\sigmav^{\alpha})}\rangle_{\{c_{ij}J_{ij}\}}
\nonumber\\
\hspace*{-20mm}
&=&
\rme^{- \rmi\theta\sum_{\alpha  i}\hat{h}_i^{\alpha}}\langle
c_{12}\delta [ h \!-\! H_2^1\!+\! 2J_{12}\tilde s]\rme^{-
\rmi\sum_{i\neq j} c_{ij}J_{ij}\sum_{\alpha}\hat{h}_i^{\alpha}
\sigma_j^{\alpha}}\rangle_{\{c_{ij}J_{ij}\}}
\\
\hspace*{-20mm}
&=&
\rme^{- \rmi\theta\sum_{\alpha  i}\hat{h}_i^{\alpha}}
\langle
c_{12}\delta [ h\! -\! H_2^1\!+\! 2J_{12}\tilde s]\rme^{-
\rmi\sum_{i< j} c_{ij}J_{ij}\sum_{\alpha}[\hat{h}_i^{\alpha}
\sigma_j^{\alpha}+ \hat{h}_j^{\alpha}
\sigma_i^{\alpha}]}\rangle_{\{c_{ij}J_{ij}\}}
\nonumber
\end{eqnarray}
We then average over the connectivity disorder $\{c_{ij}\}$, which is defined by (\ref{eq:P(c)}), followed by the bond disorder $\{J_{ij}\}$:
\begin{eqnarray}
\hspace*{-15mm}
\bra \ldots\ket_{\{c_{ij}J_{ij}\}}
&=&\frac{c}{N}\rme^{- \rmi\theta\sum_{\alpha
i}\hat{h}_i^{\alpha}}\Big\langle
\delta [ h\! -\! H_2^1\!+ \!2J_{12}\tilde
s]\rme^{- \rmi J_{12}\sum_{\alpha}[\hat{h}_1^{\alpha}
\sigma_2^{\alpha}+ \hat{h}_2^{\alpha}
\sigma_1^{\alpha}]}
\nonumber\\
\hspace*{-0mm}
&&\times\prod_{
     i<j, j\neq 2}
\Big\{\frac{c}{N}\rme^{- \rmi
J_{ij}\sum_{\alpha}[\hat{h}_i^{\alpha}
\sigma_j^{\alpha}+ \hat{h}_j^{\alpha}
\sigma_i^{\alpha}]}+ \big(1-\frac{c}{N}\big)\Big\}\Big\rangle_{\{J_{ij}\}}
\nonumber\\
\hspace*{-0mm}
&=&\frac{c}{N}\rme^{- \rmi\theta\sum_{\alpha  i}\hat{h}_i^{\alpha}}
\int\! \mathrm d J~ P(J)~\delta [ h\! -\! H_2^1\!+\! 2J\tilde
s]\rme^{- \rmi J\sum_{\alpha}[\hat{h}_1^{\alpha}
\sigma_2^{\alpha}+ \hat{h}_2^{\alpha}
\sigma_1^{\alpha}]}
\nonumber\\
\hspace*{-0mm}
&&\times\prod_{i<j,
     j\neq 2}\Big\{\frac{c}{N}\int\! \mathrm d J~ P(J)~\rme^{- \rmi
J\sum_{\alpha}[\hat{h}_i^{\alpha}
\sigma_j^{\alpha}+ \hat{h}_j^{\alpha}
\sigma_i^{\alpha}]}+ 1-\frac{c}{N}\Big\}
\end{eqnarray}
Finally, we re-exponentiate the last line of the above expression, giving
\begin{eqnarray}
\hspace*{-15mm}
\bra \ldots\ket_{\{c_{ij}J_{ij}\}}
&=&\frac{c}{N}\rme^{- \rmi\theta\sum_{\alpha i}\hat{h}_i^{\alpha}}
\int\! \mathrm d J~ P(J)~\delta [ h\! -\! H_2^1\!+\! 2J\tilde
s]\rme^{- \rmi J\sum_{\alpha}[\hat{h}_1^{\alpha}
\sigma_2^{\alpha}+ \hat{h}_2^{\alpha}
\sigma_1^{\alpha}]}\nonumber\\
&&\hspace*{-5mm}
\times\exp\Big[\frac{c}{2N}\sum_{ij}\Big[\int\! \mathrm d J~
P(J)\rme^{- \rmi J\sum_{\alpha}[\hat{h}_i^{\alpha}
\sigma_j^{\alpha}+ \hat{h}_j^{\alpha}
\sigma_i^{\alpha}]}\!- 1\Big]+ O(1)\Big]
\end{eqnarray}

\section{Calculation of the RS saddle-point equations}\label{section:RSeq}
We compute the RS versions of the  kernel (\ref{eq:A}) and the saddle-point equation (\ref{eq:D}).
Assuming replica-symmetry transforms in these equations the averages over the effective measure,  $\langle\ldots\rangle_M\rightarrow\langle\ldots\rangle_{M_{RS}}$, with the definition (\ref{eq:Mrs}).
 In (\ref{eq:A},\ref{eq:D})  this gives
\begin{eqnarray}
\hspace*{-15mm}
D(s,h)
&=&\frac{1}{M_{RS}^{n}}\sum_{k\geq 0}\frac{c^k}{k!}\rme^{-
c}\int\!\prod_{\ell=1}^k\Big\{ \mathrm d J_\ell P(J_\ell)
\left\lbrace \mathrm d P_\ell\right\rbrace W[\{P_\ell\}]\Big\}
\nonumber\\
\hspace*{-15mm}
&&\times \Big\{
\sum_{\sigma_{1}}\int\! \mathrm d
H_{1} \mathrm d
\hat{h}_{1}d(\sigma_{1},H_{1})\rme^{\rmi\hat{h}_{1}[
H_{1}-\theta]}\delta_{s,\sigma_{1}}\delta
[h- H_{1}]
\nonumber\\
\hspace*{-15mm}
&&\hspace*{10mm} \times \prod_{\ell=1}^k\Big[
\sum_{\sigma^{1}_{\ell}}\int\!\mathrm d \hat{h}^{1}_{\ell}
P_\ell(\sigma^{1}_{\ell},\hat{h}^{1}_{\ell})\rme^{- \rmi
J_\ell[\hat{h}_{1}\sigma^{1}_\ell+\hat{h}^{1}_{\ell} \sigma_{1}]}\Big]\Big\}
\nonumber\\
\hspace*{-15mm}
&&\times \prod_{\alpha=2}^n\Big\{
\sum_{\sigma_{\alpha}}\int\! \mathrm d
H_{\alpha} \mathrm d
\hat{h}_{\alpha}d(\sigma_{\alpha},H_{\alpha})\rme^{\rmi\hat{h}_{\alpha}[
H_{\alpha}-\theta]}
\nonumber\\
\hspace*{-15mm}
&&\hspace*{10mm}\times \prod_{\ell=1}^k\Big[
\sum_{\sigma^{\alpha}_{\ell}}\int\!\mathrm d \hat{h}^{\alpha}_{\ell}
P_\ell(\sigma^{\alpha}_{\ell},\hat{h}^{\alpha}_{\ell})\rme^{- \rmi
J_\ell[\hat{h}_{\alpha}\sigma^{\alpha}_\ell+\hat{h}^{\alpha}_{\ell} \sigma_{\alpha}]}\Big]\Big\}
\nonumber\\
\hspace*{-15mm}
&=&\sum_{k\geq 0}\frac{c^k}{k!}\rme^{-
c}\int\!\prod_{\ell=1}^k\Big\{ \mathrm d J_\ell P(J_\ell)
\left\lbrace \mathrm d P_\ell\right\rbrace W[\{P_\ell\}]\Big\}
\frac{Z[\{P_1,\ldots,P_k\}]^{n-1}}{M_{RS}^{n}}
\nonumber\\
\hspace*{-15mm}
&&\times \sum_{\sigma}\prod_{\ell=1}^k\Big[
\sum_{\sigma_{\ell}}\int\!\mathrm d \hat{h}_{\ell}
P_\ell(\sigma_{\ell},\hat{h}_{\ell})\rme^{- \rmi J_\ell\hat{h}_{\ell}
\sigma}\Big]\nonumber
\\
\hspace*{-15mm} &&\times d\big(\sigma,\sum_{\ell=1}^k
J_\ell\sigma_{\ell}\!+\!\theta\big)\delta_{s,\sigma_{}}\delta
[h\!-\!\sum_{\ell=1}^k
J_\ell\sigma_{\ell}\!-\!\theta]
\nonumber\\
\hspace*{-15mm}
&=& \sum_{k\geq 0}\frac{c^k}{k!}\rme^{-
c}\int\!\prod_{\ell=1}^k\Big\{ \mathrm d J_\ell P(J_\ell)
\left\lbrace \mathrm d P_\ell\right\rbrace W[\{P_\ell\}]\Big\}
\frac{Z[\{P_1,\ldots,P_k\}]^{n-1}}{M_{RS}^{n}}
\nonumber\\
\hspace*{-15mm}
&&\times d(s,h) \prod_{\ell=1}^k\Big[
\sum_{\sigma_{\ell}}\int\!\mathrm d \hat{h}_{\ell}
P_\ell(\sigma_{\ell},\hat{h}_{\ell})\rme^{- \rmi J_\ell\hat{h}_{\ell}
s}\Big] \delta
[h\!-\!\sum_{\ell=1}^k
J_\ell\sigma_{\ell}\!-\!\theta]
\label{eq:DbeforeReplLimit}
\end{eqnarray}
\clearpage
and
\begin{eqnarray}
A[s,s^\prime;h,
h^\prime;\tilde s]&=& \frac{\tilde{A}[s,s^\prime;h,h^\prime;\tilde{s}]}
{\sum_{\sigma\sigma^\prime}\int\!dH dH^\prime~\tilde{A}[\sigma,\sigma^\prime;H,H^\prime;\tilde{s}]}
\label{eq:AbeforeReplLimit}
\end{eqnarray}
where
\begin{eqnarray}
\hspace*{-20mm}
\tilde{A}[s,s^\prime;h,
h^\prime;\tilde s]
&=&
\langle\delta_{{s^\prime},\sigma_1}\delta_{s, {\sigma}^\prime_1}\delta
[h^\prime\! -\! H_1]\delta [ h\! -\! H^\prime_1\!+\! 2J\tilde s]\rme^{- \rmi J[\hat{\hv}
\cdot{\sigmav^\prime}+\hat{\hv}^\prime\cdot
\sigmav]}\rangle_{J,M_{RS},M^\prime_{RS}}
\nonumber
\\
\hspace*{-20mm}
&&\hspace*{-15mm}
=
 \frac{1}{M_{RS}^{2n}}\int\! \mathrm d J P(J)\sum_{\sigmav {\sigmav^\prime}}\int\! \mathrm d \Hv \mathrm d\Hv^\prime
\mathrm d \hat{\hv}\mathrm d \hat{\hv}^\prime \delta_{{s^\prime},\sigma_1}\delta_{s,\sigma^\prime_1}\delta
[h^\prime\! -\! H_1]\delta [ h\! -\! H^\prime_1\!+\! 2J\tilde s]
\nonumber\\
\hspace*{-20mm}
&&\hspace*{-10mm} \times\sum_{k\geq 0}\frac{c^k}{k!}\rme^{- c}\int\!\prod_{\ell=1}^k\Big\{ \mathrm d J_\ell P(J_\ell)
\left\lbrace \mathrm d P_\ell\right\rbrace W[\{P_\ell\}]\Big\}
\nonumber\\
\hspace*{-20mm}
&&\hspace*{-10mm} \times \prod_{\alpha=1}^n d(\sigma_{\alpha},H_{\alpha})\rme^{\rmi\hat{h}_{\alpha}[
H_{\alpha}- \theta- J\sigma^\prime_{\alpha}]}\prod_{\ell=1}^k\Big[ \sum_{\sigma^{\alpha}_{\ell}}\int\!\mathrm d
\hat{h}^{\alpha}_{\ell} P_\ell(\sigma^{\alpha}_{\ell},\hat{h}^{\alpha}_{\ell})\rme^{- \rmi J_\ell[\hat{h}_{\alpha}\sigma^{\alpha}_\ell+\hat{h}^{\alpha}_{\ell} \sigma_{\alpha}]}
\Big] \hspace*{-10mm}
\nonumber\\
\hspace*{-20mm}
&&\hspace*{-10mm}
\times \sum_{m\geq 0}\frac{c^m}{m!}\rme^{- c}\int\!\prod_{r=1}^m\Big\{ \mathrm d J_r P(J_r)
\left\lbrace \mathrm d Q_r\right\rbrace W[\{Q_r\}]\Big\}
\nonumber \\
\hspace*{-20mm}
&&\hspace*{-10mm} \times \prod_{\alpha=1}^n d(\sigma_{\alpha}^\prime,H^\prime_{\alpha})\rme^{\rmi\hat{h}^\prime_{\alpha}[
H^\prime_{\alpha}- \theta- J \sigma_{\alpha}]}\prod_{r=1}^m\Big[ \sum_{\sigma^{\alpha}_{r}}\int\!\mathrm d
\hat{h}^{\alpha}_{r} Q_r(\sigma^{\alpha}_{r},\hat{h}^{\alpha}_{r})\rme^{- \rmi J_r[\hat{h}^\prime_{\alpha}\sigma^{\alpha}_r+\hat{h}^{\alpha}_{r} \sigma^\prime_{\alpha}]}\Big]
\hspace*{-10mm}
\nonumber\\
\hspace*{-20mm}
&&\hspace*{-15mm}
=
\frac{1}{M_{RS}^{2n}}\sum_{k\geq 0}\frac{c^k}{k!}\rme^{- c}\int\!\prod_{\ell=1}^k\Big\{
\mathrm d J_\ell P(J_\ell)
\left\lbrace \mathrm d P_\ell\right\rbrace W[\{P_\ell\}]\Big\}
\nonumber\\
&&\hspace*{-10mm} \times \sum_{m\geq 0}\frac{c^m}{m!}\rme^{- c}\int\!\prod_{r=1}^m\Big\{ d J_r P(J_r)
\left\lbrace \mathrm d Q_r\right\rbrace W[\{Q_r\}]\Big\}
\int\!dJ~P(J)
\nonumber \\
\hspace*{-20mm}
&&\hspace*{-10mm}
\times
\sum_{\sigma \sigma^\prime}\prod_{\ell=1}^k\Big[
\sum_{\sigma_{\ell}}\int\!\mathrm d \hat{h}_{\ell}
P_\ell(\sigma_{\ell},\hat{h}_{\ell})\rme^{- \rmi J_\ell \hat{h}_{\ell} \sigma }\Big]
d\big(\sigma,\sum_{\ell=1}^k
J_\ell\sigma_{\ell}\!+\! \theta\!+\! J\sigma^\prime\big)
\nonumber\\
\hspace*{-20mm}
&&\hspace*{-10mm} \times \prod_{r=1}^m\Big[
\sum_{\sigma_{r}}\int\!\mathrm d \hat{h}_{r}
Q_r(\sigma_{r},\hat{h}_{r})\rme^{- \rmi
J_r\hat{h}_{r} \sigma^\prime}\Big] d\big(\sigma^\prime,\sum_{r=1}^m
J_r\sigma_{r}\!+\! \theta\!+\! J \sigma\big)
\nonumber\\
\hspace*{-20mm}
&&\hspace*{-10mm}\times
\delta_{s,\sigma^\prime}\delta [ h\! -\! \sum_{r=1}^m
J_r\sigma_{r}\!-\! \theta\!-\! J \sigma\!+\! 2J\tilde s]\delta_{{s^\prime},\sigma}\delta
[h^\prime \!-\! \sum_{\ell=1}^k
J_\ell\sigma_{\ell}\!-\! \theta\!-\! J\sigma^\prime]
\end{eqnarray}
In the replica limit $n\rightarrow0$ we get $\lim_{n\to 0}M_{RS}^{n}=\lim_{n\to 0}Z[\{P_1,\ldots,P_k\}]^n=1$, and also
$\lim_{n\to 0} \sum_{\sigma\sigma^\prime}\int\!dH dH^\prime~\tilde{A}[\sigma,\sigma^\prime;H,H^\prime;\tilde{s}]=1$.
As a result, the above expressions (\ref{eq:DbeforeReplLimit},\ref{eq:AbeforeReplLimit})
reduce to equations (\ref{eq:Drs1},\ref{eq:Ars1}).
\section{Reduction of PDE to the system of ODEs}\label{section:ODE}
In this section we show how the diffusion equation (\ref{eq:diffusion}), written in terms of the kernels (\ref{eq:VCjsfield},\ref{def:Avc}), can be reduced to a system of ordinary differential equations.
 The discrete nature of the fields (\ref{eq:VCfield}) allows to write (\ref{eq:VCjsfield},\ref{def:Avc}) in terms
 of the probability distributions (\ref{def:Psn}) and (\ref{def:Asn}). Inserting (\ref{def:Psn}) and (\ref{def:Asn})
 into  both sides of (\ref{eq:diffusion}) gives
\begin{eqnarray}
\hspace*{-20mm}
\frac{\partial}{\partial t}\sum_{n\geq0}P_t(s,n)\delta (h\!-\! Jn\!- \theta) &=&
\frac{1}{2}\left [1\!+\! s\tanh[\beta h]\right
]\sum_{n\geq0}P_t(\!-\! s,n)\delta (h\!-\! Jn\!-\! \theta)
\nonumber\\
\hspace*{-20mm}
&&
-\frac{1}{2}\left [1\!-\! s\tanh[\beta h]\right
]\sum_{n\geq0}P_t(s,n)\delta (h\!-\! Jn\!-\! \theta)\nonumber
\\
\hspace*{-20mm}
&&+ \frac{1}{2}c\sum_{{s^\prime}}\int\!\mathrm{d}{h^\prime}~
[1\!-\! {s^\prime}\tanh[\beta {h^\prime}]]
\sum_{n n^\prime\geq 0}A_t[s,s^\prime;n,n^\prime]
\nonumber
\\
\hspace*{-20mm}
&&\times \delta
[h^\prime\!\! -\! J(n^\prime\!+\!\delta_{s,- 1})\!-\! \theta]
\delta [ h \!-\! J(n\!+\!\delta_{s^\prime,- 1}\!+\!s^\prime)\!-\! \theta]
\nonumber\\
\hspace*{-20mm}
&&-\frac{1}{2}c\sum_{{s^\prime}}\int\!\mathrm{d}{h^\prime}~
[1\!-\! {s^\prime}\tanh[\beta {h^\prime}]]
\sum_{n n^\prime\geq 0}A_t[s,s^\prime;n,n^\prime]\nonumber
\\
\hspace*{-20mm}
&&\times \delta
[h^\prime\!\! -\! J(n^\prime\!+\!\delta_{s,-
1})\!-\! \theta]\delta [ h\! -\! J(n\!+\!\delta_{s^\prime,-
1})\!-\! \theta]
\end{eqnarray}
In the left-hand side we move the time derivative inside the summation, and in the right-hand side we
sum over $s^\prime$ and integrate over $h^\prime$. This leads to
\begin{eqnarray}
\hspace*{-20mm}
\sum_{n\geq0}\frac{\mathrm d}{\mathrm d t}P_t(s,n)\delta (h\!- \!Jn\!-\! \theta)
&=& \sum_{n\geq0}\delta (h\!-\! Jn\!-\! \theta)\Big\{
\frac{1}{2}\left [1\!+\! s\tanh[\beta (Jn\!+\!
\theta)]\right
]P_t(\!-\! s,n)
\nonumber\\
\hspace*{-20mm}
&&\hspace*{10mm}- \frac{1}{2}\left [1\!-\! s\tanh[\beta (Jn\!+\!
\theta)]\right]P_t(s,n)
\Big\}
\nonumber\\
\hspace*{-20mm}
&&+\frac{1}{2}c\sum_{n\geq 0} \delta (h \!-\! Jn\!-\! \theta)\sum_{n^\prime\geq 0}\Big\{
\nonumber
\\
\hspace*{-20mm}&&
\hspace*{10mm}
[1\!+\! \tanh[\beta (Jn^\prime \!+\! \theta\!+\! J \delta_{s,- 1})]]A_t[s,\!-\!1;n,n^\prime]\nonumber\\
& &\hspace*{10mm}-
[1\!-\! \tanh[\beta( Jn^\prime \!+\! \theta\!+\! J \delta_{s,- 1})]]A_t[s,1;n,n^\prime]\Big\}
\nonumber\\
\hspace*{-20mm}
&&+\frac{1}{2}c\sum_{n\geq 1}\delta (h \!-\! Jn\!-\! \theta)\sum_{n^\prime\geq 0}\Big\{
\nonumber\\ \hspace*{-20mm}&&
\hspace*{10mm}
[1\!-\! \tanh[\beta( Jn^\prime \!+\! \theta\!+\! J \delta_{s,- 1})]]A_t[s,1;n\!-\! 1,n^\prime]
\nonumber\\
&&\hspace*{-10mm}-
[1\!+\! \tanh[\beta (Jn^\prime\!+\!\theta\!+\! J \delta_{s,- 1})]]A_t[s,\!-\! 1;n\!-\! 1,n^\prime]\Big\}
\end{eqnarray}
It follows from the above that the evolution in time of $P_t(s,n)$ is governed by (\ref{eq:ODE}).
\section{Initial conditions}\label{section:Initial conditions}
In this section we compute the values of the probability distribution $P(s,n)$, the functional distribution $\tilde W[\{\hat P\}]$ and the function $d(s,Jn+ \theta)$ at time $t=0$. We choose an initial state of the system in which all individual spin values are drawn randomly and independently,
according to
\begin{eqnarray}
P_0(\sigmav) &=& \prod_{i=1}^N\lbrace\frac{1}{2}(1+ m_0)\delta_{\sigma_i;1}+ \frac{1}{2}(1- m_0)\delta_{\sigma_i;- 1}\rbrace\label{eq:Prob0}
\end{eqnarray}
where $m_0\in[-1,1]$ is the prescribed initial magnetization. It follows that the joint spin-field distribution
(\ref{eq:VCjsfield}) at $t=0$ is given by
\begin{eqnarray}
\hspace*{-15mm}
D_0(s,h) &=& \lim_{N\rightarrow\infty}\sum_{\sigmav}P_0(\sigmav)\frac{1}{N}\sum_i^N\delta_{s,\sigma_i}\langle \delta
[h- h_i(\sigmav)]\rangle_{\{c_{ij}\}}
\nonumber \\
\hspace*{-15mm}
&=& \frac{1}{2}(1\!+\! s m_0)\sum_{k\geq 0}\frac{c^k}{k!}\rme^{- c}
\prod_{\ell=1}^k\Big\{\sum_{\sigma_{\ell}}\frac{1}{2}(1\!+\! \sigma_\ell m_0)\Big\}\delta
[h\!- \!J\sum_{\ell=1}^k \delta_{\sigma_{\ell},- 1}\!-\! \theta]\nonumber\\
\hspace*{-15mm}
&=&\frac{1}{2}(1\!+\! s m_0)\sum_{n\geq 0}\frac{[\frac{1}{2}c(1\!-\! m_0)]^n}{n!}\exp\Big[\!-\! \frac{1}{2}c(1\!-\! m_0)\Big]
\delta[h\!- \!Jn\!-\! \theta]
\label{eq:D0}
\end{eqnarray}
The initial conditions for the system (\ref{eq:ODE}) follow directly from the above expression:
\begin{eqnarray}
P_0(s,n)&=&\frac{1}{2}(1+ s m_0)\frac{[\frac{1}{2}c(1- m_0)]^n}{n!}\exp[- \frac{1}{2}c(1- m_0)]\label{eq:P0(s,n)}
\end{eqnarray}
Furthermore, we see that the joint spin-field distribution (\ref{eq:D0}) indeed takes the desired form of the saddle-point equation
(\ref{eq:DvcRS2}), with the functional distribution
 \begin{eqnarray}
 \tilde W[\{\hat P\}]=\prod_{\sigma,\acute{\sigma}}\delta[\hat P(\sigma\vert J\delta_{\acute\sigma;- 1})- \frac{1}{2}(1+ \sigma m_0)]\label{eq:W0}
 \end{eqnarray}
 and with
 \begin{eqnarray}d(s,Jn+ \theta)=\frac{1}{2}(1+ s m_0)\label{eq:d0}\end{eqnarray}
It is a trivial matter to show that (\ref{eq:W0},\ref{eq:d0}) are indeed
 the solutions of equation (\ref{eq:WvcFT}).

\section{Population dynamics}\label{section:population}
The functional saddle-point equations (\ref{eq:WvcFT},\ref{eq:Psn}) cannot in general general be solved analytically (one trivial exception is
the infinite temperature regime). We therefore resort to the so-called population dynamics algorithm \cite{BetheSG} to obtain solutions numerically,
solving equations (\ref{eq:WvcFT}) and (\ref{eq:Psn}) simultaneously for the functional distribution $\tilde W[\{\hat P\}]$ and the function $d(s,Jn+ \theta)$, given the (known) values of the probability distribution $P_t(s,n)$ at time $t$. We create a population of $\mathcal{N}$ $2\!\times\!2$ matrices  $\hat P_i(\sigma \vert J\delta_{\sigma^\prime,- 1})$, where $i=1\ldots \mathcal{N}$, and we initialize the numbers
$d(s,Jn+ \theta)$, where $s\in\{-1,1\}$ and $n\in\{0,1,2,\ldots\}$. We then execute an iterative process, whereby at each step we update the population of matrices and the numbers $d(s,Jn+ \theta)$ as follows:
\begin{enumerate}
  \item a number $k$ is drawn from the Poisson distribution $P_c(k)$ (\ref{eq:P(c)})
  \item $k$ members $\hat{P}_i(\sigma \vert J\delta_{\sigma^\prime,- 1})$ are selected randomly and independently from the population
  \item a new value for $P(\sigma \vert J\delta_{\sigma^\prime,- 1})$ is calculated according to
\begin{eqnarray}
\hspace*{-30mm}
\hat P_{\rm new}(\sigma \vert J\delta_{\sigma^\prime,- 1})=\frac{\prod_{\ell=1}^k\Big[ \sum_{\sigma_{\ell}}
\hat P_\ell(\sigma_{\ell}\vert J\delta_{\sigma, - 1})\Big] d(\sigma,J\sum_{\ell=1}^k \delta_{\sigma_\ell ,- 1}\!+\! \theta\!+\! J\delta_{\sigma^\prime,- 1})}{\sum_{\sigma}\prod_{\ell=1}^k\Big[ \sum_{\sigma_{\ell}}
\hat P_\ell(\sigma_{\ell}\vert J\delta_{\sigma, - 1})\Big] d(\sigma,J\sum_{\ell=1}^k \delta_{\sigma_\ell ,- 1}\!+\! \theta)}\label{eq:Pupdate}~\end{eqnarray}
  \item a member of the population is selected randomly, and replaced with the newly computed value $\hat P_{\rm new}(\sigma \vert J\delta_{\sigma^\prime,- 1})$
  \item a new function $d(s,Jn+ \theta)$ is computed according to
\begin{eqnarray}
\hspace*{-25mm}
d_{\rm new}(s,Jn+ \theta)&=&P_t(s,n)\times\Bigg[\sum_{k\geq 0}\frac{c^k}{k!}\rme^{- c}\int\prod_{\ell=1}^k\Big\{
\{\mathrm d \hat P_\ell\} \tilde W[\{\hat P_\ell\}]\Big\}\label{eq:dUpdate} \\
\hspace*{-20mm}
&&\times \frac{\prod_{\ell=1}^k\Big[ \sum_{\sigma_{\ell}}
\hat P_\ell(\sigma_{\ell}\vert J \delta_{s,- 1})\Big]\delta_{n,\sum_{\ell=1}^k \delta_{\sigma_{\ell},- 1}}}
{\sum_{\sigma}\prod_{\ell=1}^k\Big[\sum_{\sigma_{\ell}}
\hat P_\ell(\sigma_{\ell}\vert J \delta_{\sigma,- 1})\Big] d\big(\sigma,J\sum_{\ell=1}^k \delta_{\sigma_{\ell},- 1}+ \theta\big)}\Bigg]^{- 1}\nonumber
\end{eqnarray}
\end{enumerate}
Here averaging over the functional measure $\tilde{W}$ is defined as averaging over the {\em actual} instantaneous population of $2\!\times\!2$ matrices. This iteration is repeated until the values of the function $d(s,Jn+ \theta)$ and the statistical properties of the population are stationary. The population measure $\tilde{W}$ will now be an estimate of the functional distribution (\ref{eq:WvcFT}), and the function $d(s,Jn+ \theta)$ is a fixed point of the iteration equation (\ref{eq:dUpdate}), i.e. a solution of our original saddle-point equation.

\end{document}